\documentclass[12pt]{article}
\usepackage{amsmath,amssymb,amsthm}
\title{Symmetries of Discrete Dynamical Systems Involving Two Species}
\vspace{-0.5cm}
\author{\\
\large{D. \textsc{ G\'omez-Ullate}}\footnote{dgu@eucmos.sim.ucm.es} \\
\small
\begin{tabular}{c}
Departamento de F\'{\i}sica Te\'orica II\\
Facultad de Ciencias F\'{\i}sicas\\
28040 Universidad Complutense\\
Madrid, Spain
\end{tabular}\\
\large{S. \textsc{Lafortune}}\footnote{lafortus@crm.umontreal.ca}
 \large{and P. \textsc{Winternitz} }\footnote{wintern@crm.umontreal.ca} \\
\small
\begin{tabular}{c}
 Centre de Recherches Math{\'e}matiques\\
 Universit{\'e} de Montr{\'e}al,\\
 C. P. 6128, Succ.~Centre-ville,\\ 
Montr{\'e}al (QC), H3C 3J7, Canada\\
\end{tabular}\\
}
\numberwithin{equation}{section}

\begin{document}
\maketitle
\def\Ima{\rm Im}
\def\Re{\rm Re}
\def\xid{\xi_{n-1}}
\def\xiu{\xi_{n+1}}
\def\xin{\xi_n}
\def\ap{a_{1,n}}
\def\apu{a_{1,n+1}}
\def\apd{a_{1,n-1}}
\def\as{a_{2,n}}
\def\asu{a_{2,n+1}}
\def\asd{a_{2,n-1}}
\def\u{u_n}
\def\v{v_n}
\def\a{a_n}
\def\b{b_n}
\def\c{c_n}
\def\d{d_n}
\def\vu{v_{n+1}}
\def\vd{v_{n-1}}
\def\au{a_{n+1}}
\def\ad{a_{n-1}}
\def\cu{c_{n+1}}
\def\cd{c_{n-1}}
\def\bu{b_{n+1}}
\def\bd{b_{n-1}}
\def\du{d_{n+1}}
\def\dd{d_{n-1}}
\def\uu{u_{n+1}}
\def\ud{u_{n-1}}
\newcommand{\ts}{\textstyle}
\newcommand{\ds}{\displaystyle}
\newcommand{\dt}{\partial_{t}}
\newcommand{\duk}{\partial_{u_{k}}}
\newcommand{\dvk}{\partial_{v_{k}}}
\newcommand{\dun}{\partial_{u_{n}}}
\newcommand{\dvn}{\partial_{v_{n}}}
\newcommand{\vect}[2]{\left( \begin{array}{c} #1 \\#2 \end{array} \right)}
\newtheorem{theorem}{Theorem}

\newcommand{\mat}[4]{\ensuremath \left(  \begin{array}{cc} #1 & #2 \\ 
#3 & #4 \end{array} \right) }
\vspace{-1cm}
\abstract{The Lie point symmetries of a coupled system
of two nonlinear differential-difference equations are 
investigated. It is shown that in special cases the symmetry group
can be infinite dimensional, in other cases up to \mbox{$10$ dimensional}.
The equations can describe the interaction of two long molecular chains,
each involving one type of atoms.}
\markboth{}{Symmetries of Discrete Dynamical Systems}

{\noindent}
\newpage
\section{INTRODUCTION}

The purpose of this article is to perform a symmetry analysis of a system of 
two coupled differential-difference equations of the form 
\begin{equation}\label{system}
\begin{array}{c}
\ds{E_1=\ddot{u}_n-F_n(t,\ud,\u,\uu,\vd,\v,\vu)=0,} \\
\ds{E_2=\ddot{v}_n-G_n(t,\ud,\u,\uu,\vd,\v,\vu)=0.}
\end{array}
\end{equation}
The dots denote time derivatives. The discrete variable $n$ plays the role 
of a space variable; it labels positions along a one dimensional lattice. 
The functions $F_n$ and $G_n$ represent interactions e.g. between different 
atoms along a double chain of molecules. The functions $F_n$ and
$G_n$ are a priori unspecified; our aim is to classify equations of the type
(\ref{system}) according to the Lie point symmetries that they allow. 
The interactions in such a model depend on up to six neighbouring particles.
For instance, we can interpret $\u$ and $\v$ as deviations from equilibrium
positions of two different types of atoms, say type $U$ and type $V$. The
 accelerations 
$\ddot{u}_n$ and $\ddot{v}_n$ depend on the deviations $u$ and $v$ of both 
types of atoms at the neighbouring sites $n-1$, $n$ and $n+1$. We do not 
restrict
to two-body forces, nor do we impose translational invariance for the chain.
We do however assume there is no dissipation, i.e. system (\ref{system})
does not involve first derivatives with respect to time.

Such differential-difference equations typically arise when modeling 
phenomena
in molecular physics, biophysics, or simply coupled oscillations in
classical mechanics [1,2,3].

A recent article [4] was devoted to a similar problem, but was concerned 
with a single species, i.e. one dependent variable $u_n(t)$. The approach
adopted here is similar to that of Ref.4. Thus, we shall consider only
symmetries acting on the continuous variables $t$, $u_n$ and $v_n$. 
Transformations of the discrete variable $n$ must then be studied
separately.

Several different treatments of Lie symmetries of difference 
and differential-difference equations exist in the literature [4-13].
The one adopted in this article is that of Ref.4,5,6. It has been
called the ``intrinsic method'', makes use of a Lie algebraic approach
and is entirely algorithmic. The Lie algebra of the symmetry group, the
``symmetry algebra'' for short, is realized by vector fields
of the form
\begin{equation}\label{vfield}
\hat{X}=\tau(t,\u,\v)\dt+\phi_n(t,\u,\v)\dun+\psi_n(t,\u,\v)\dvn.
\end{equation}
The algorithm for finding the functions $\tau$, $\phi_n$ and $\psi_n$ in 
(\ref{vfield}) is to construct the appropriate prolongation $pr\hat{X}$
of $\hat{X}$ (see Ref.$4$ ,$5$, $6$ and Section \ref{s2} below) and to impose that
it should annihilate the studied system of equations on their solution
set
\begin{equation}\label{invariance}
\mbox{pr}\hat{X}E_1|_{E_1=E_2=0}=0,\;\;
\mbox{pr}\hat{X}E_2|_{E_1=E_2=0}=0.
\end{equation}

Our first step is to find and classify all interactions ($F_n$, $G_n$) 
for which the system (\ref{system}) allows at least a one dimensional 
symmetry algebra. The next step is to specify the interactions further 
and to find all those that allow a higher dimensional, possibly 
infinite dimensional, symmetry algebra.

As in previous articles [4,14] our classification will be up to conjugacy
under a group of ``allowed transformations''. 
These are fiber preserving locally 
invertible point transformations
\begin{equation}\label{all. transf.}
\u=\Omega_n(\tilde{u}_n,\tilde{v}_n,\tilde{t}),\;\;
\v=\Gamma_n(\tilde{u}_n,\tilde{v}_n,\tilde{t}),\;\;t=t(\tilde{t})
\end{equation}
that preserve the form of equations (\ref{system}), but not necessarily
the functions $F_n$ and $G_n$ (they go into new functions $\tilde{F}_n$
and $\tilde{G}_n$ of the new arguments).

Throughout the article we assume that both $F_n$ and $G_n$ depend on at 
least 
one of the quantites $\ud$, $\uu$, $\vd$, $\vu$, so that nearest neighbours
are genuinely involved. In the bulk of the article the interaction is assumed
to be nonlinear.

In Section \ref{s2} we formulate the problem, establish the general form of the 
elements of the symmetry algebra and present the determining equations 
for the symmetries. We also derive the ``allowed transformations'' 
under which
we classify the interactions and their symmetries. Section \ref{s3}
 is devoted to
a classification of interactions $F_n$, $G_n$ allowing at least 
a one dimensional symmetry algebra. Ten classes of such interactions exist, 
each involving $2$ arbitrary functions of $6$ variables. In Section \ref{s4}
 we study higher dimensional symmetry algebras and introduce an 
important restriction. We first prove that $4$ equivalence classes
of symmetry algebras isomorphic to ${\mbox sl}(2,\mathbb{R})$ exist. 
Then we restrict to just one of them, ${\mbox sl}(2,\mathbb{R})_1$ 
generating a gauge group acting only on the fields $\u$ and $\v$
(in a global, coordinate independent manner). We describe
all symmetry algebras, containing the chosen ${\mbox sl}(2,\mathbb{R})$ as a
subalgebra. In Section \ref{s5} we obtain the invariant interactions for 
all algebras containing ${\mbox sl}(2,\mathbb{R})_1$. The results are 
summed up and discussed in Section \ref{s6} where we also outline future 
work to be done.


\section{FORMULATION OF THE PROBLEM}\label{s2}
To find the Lie point symmetries of the system (\ref{system}) we write the
second prolongation of the vector field (\ref{vfield}) in the form 
\cite{LW96,LW91,LW93}:

\begin{equation} \label{prolongation}
\begin{aligned}
\mbox{pr}^{(2)}\hat X& =  \tau(t,u_{n},v_{n})\dt + \sum_{k=n-1}^{n+1} 
\phi_{k}(t,u_{n},v_{n})\duk  \\
&\quad + \sum_{k=n-1}^{n+1} 
\psi_{k}(t,u_{n},v_{n})\dvk
+ \phi_{n}^{tt} \partial_{\ddot{u}_n} + \psi_{n}^{tt}\partial_{\ddot{u}_n},
\end{aligned}
\end{equation}
with
\begin{equation}
\begin{array}{c}
\phi_{n}^{tt}=D_{t}^{2}\phi_{n} -(D_{t}^{2}\tau)\dot u_{n} -
2(D_{t}\tau)\ddot u_{n}, \\[2mm]
\psi_{n}^{tt}=D_{t}^{2}\psi_{n} -(D_{t}^{2}\tau)\dot v_{n} -
2(D_{t}\tau)\ddot v_{n},
\end{array}
\end{equation}
where $D_{t}$ is the total time derivative. 
The determining equations for 
the symmetries are obtained by requiring that eq.(\ref{invariance}) be
satisfied. The obtained equations will involve terms like $\dot 
u^{k}\,,\,\dot v^{k}\,$ and $\,\dot u^{k} \dot v^{l}$. The coefficients of
each 
linearly independent term must vanish and this provides 16 linear 
differential equations that are easy to solve and do not involve the 
interaction functions $F_{n}\,,G_{n}$. The result is that an element $\hat
X$ of the symmetry algebra must have the form
\begin{equation}\label{symm.generator}
\begin{aligned}
\hat X = \tau(t)\dt &+ \left[\left(\frac{\dot \tau}{2} +a_{n}\right)u_{n}
+ b_{n}v_{n} +\lambda_{n}(t)\right]\dun \\[3mm]
&+ \left[c_{n}u_{n} + \left(\frac{\dot 
\tau}{2} +d_{n}\right)v_{n} + \mu_{n}(t)\right]\dvn,
\end{aligned}
\end{equation}
where the dots denote time derivatives.
The functions
$\tau(t)$, $\lambda_{n}(t)$, $\mu_{n}(t)$, $a_{n}$, $b_{n}$, $c_{n}$ and  
$d_{n}$ satisfy the two remaining determining equations, namely


\begin{equation}\label{remaining1}
\begin{aligned}
\frac{\ds \dddot{\tau}}{2}&\,u_{n}+\ddot \lambda_{n}+ \left( a_{n} - 
\frac{3}{2}
\dot{\tau} \right)F_{n} +b_{n} G_{n} - \tau  F_{n,t}\\
&\quad -\sum_{k=n-1}^{n+1}  F_{n,u_{k}}
\left[\left(\frac{\dot \tau}{2} +a_{k}\right)u_{k} + b_{k}v_{k} 
+\lambda_{k}(t)\right] \\
&\quad -\sum_{k=n-1}^{n+1} F_{n,v_{k}} \left[ \left( \frac{\dot \tau}{2}
+d_{k} \right)v_{k} + c_{k}u_{k} +\mu_{k}(t)\right]  =0,  \vrule width 0pt
depth 20pt
\end{aligned}
\end{equation}


\begin{equation}\label{remaining2}
\begin{aligned}
\frac{\ds \dddot{\tau}}{2}&\,v_{n}+\ddot \mu_{n}+ \left( d_{n} - 
\frac{3}{2} 
\dot{\tau} \right)G_{n} +c_{n} F_{n} - \tau  G_{n,t}\\
 &\quad -\sum_{k=n-1}^{n+1}  G_{n,u_{k}}
\left[\left(\frac{\dot \tau}{2} +a_{k}\right)u_{k} + b_{k}v_{k} 
+\lambda_{k}(t)\right] \\
&\quad -\sum_{k=n-1}^{n+1} G_{n,v_{k}} \left[ \left( \frac{\dot \tau}{2}
+d_{k} \right)v_{k} + c_{k}u_{k} +\mu_{k}(t)\right]  =0.  \vrule width 0pt
depth 20pt
\end{aligned}
\end{equation}

In eq.(\ref{symm.generator}), (\ref{remaining1}) and (\ref{remaining2}) 
the quantities $a_{n},b_{n},c_{n}$ and $d_{n}$ are independent of $t$.
To proceed further one could specify the interaction functions $F_{n}$ and
$G_{n}$. Instead, we shall assume that at least one symmetry generator 
(\ref{symm.generator}) exists and make use of allowed transformations to
simplify this vector. The second step is to find interactions $F_{n}$ and
$G_{n}$ compatible with such a symmetry.

Substituting (\ref{all. transf.}) into eq.(\ref{system}) and requiring
that 
the form of these two equations be preserved, we find that the 
allowed 
transformations are quite restricted, namely
\begin{equation}\label{allowed}
\begin{array}{c}
\displaystyle{
\vect{u_{n}(t)}{v_{n}(t)} = \mat{Q_{n}}{R_{n}}{S_{n}}{T_{n}}{\dot{\tilde
t}}^{-1/2}\,\vect{\tilde{u}_{n}(\tilde t)}{\tilde{v}_{n}(\tilde t)} + 
\vect{\alpha_{n}(t)}{\beta_{n}(t)},}
\\  \\ 
\displaystyle{\tilde t = \tilde t(t)\;\;\;,\;\; \ds\frac{d\tilde t}
{dt} \neq 0.}
\end{array}
\end{equation}
The entries $Q_n$, $R_n$, $S_n$ and $T_n$ are independant of $t$; 
$\tilde{t}(t)$
is an arbitrary locally invertible function of $t$; $\alpha_n$, 
$\beta_n$
are arbitrary functions of $n$ and $t$, and the matrix
\begin{equation}
M_n=
\left(
\begin{array}{ll}
Q_n & R_n \\ S_n & T_n
\end{array}
\right),\;\;
\det{M_n}\neq 0
\end{equation}
is nonsingular.

It will be convenient to use a short-hand notation for the vector field
$X_{n}$ of eq.(\ref{symm.generator}), namely
\begin{equation}
\left \{ \tau(t), A_{n}, \vect{\lambda_{n}(t)}{\mu_{n}(t)} \right \} 
\;\;,\;\;A_{n}= \mat{a_{n}}{b_{n}}{c_{n}}{d_{n}}.
\end{equation}

If we perform an allowed transformation (\ref{allowed}),
 then eq.(\ref{system}) goes into an equation of the same form, with
$F_{n}$ 
and $G_{n}$ replaced by
\begin{equation} 
\begin{aligned}
\vect{\tilde{F}_{n}}{\tilde{G}_{n}}& = {\dot{\tilde t}}^{-3/2}\,
M_{n}^{-1}
 \left[ \vect{F_{n}}{G_{n}} - \vect{\ddot \alpha_{n}}{\ddot\beta_{n}}
\right]\\[2mm]
&\qquad  + \left( \frac{1}{2}\,\dddot{\tilde t}\,\; {\dot{\tilde t}}^{-3}
-\frac{3}{4}\,
\ddot{\tilde t}^{\,2}\;{\dot{\tilde t}}^{-4}   
\right) \vect{\tilde{u}_{n}}{\tilde{v}_{n}},
\end{aligned}
\end{equation}
where $\tilde{F}_{n}$ and $\tilde{G}_{n}$ are functions of the new 
variables.

The vector field characterized by the triplet (\ref{symm.generator})
goes into
a new one of the same form
\begin{equation}\label{changedvf}
\left \{
\tilde{\tau}(\tilde{t}), 
 \tilde{A}_n,
\vect{\tilde{\lambda}_n(\tilde{t})}{ \tilde{\mu}_n(\tilde{t})} 
\right\} 
\end{equation}
with
\begin{equation}\nonumber
\begin{array}{l}   
\tilde{\tau}(\tilde t) = \tau(t(\tilde{t}))\dot{\tilde{t}},
 \\ [1.5mm]
\tilde{A}_n = M_{n}^{-1}A_{n}M_n\;\; \\[1.5mm]
\vect{\tilde{\lambda}_n(\tilde t)}{\tilde{\mu}_n(\tilde t)} = 
M_{n}^{-1}\dot{\tilde{t}}^{\,1/2}
\left[(A_n+\frac{\ds\dot{\tau}}{\ds 2})
\vect{\alpha_n}{\beta_n}
 -\tau\vect{\dot{\alpha}_n}{\dot{\beta}_n} 
+\vect{\lambda_n}{ \mu_n}\right].
\end{array}
\end{equation}
We shall use the allowed transformations to simplify the 
vector field, rather than the equation itself.


\section{SYSTEMS WITH ONE-DIMENSIONAL SYMMETRY GROUPS}\label{s3}

Let us now assume that the system~(\ref{system}) has at least a 
one dimensional symmetry group, generated by a vector field of the 
type 
(\ref{symm.generator}). Using allowed transformations (\ref{allowed}), we 
take $\hat X$ into one of 10 inequivalent classes.

Indeed, for $\tau \neq 0$ we can choose the function $\tilde t(t)$ so as to 
transform $\tau(t)$ into $\tau = 1$, the functions $\alpha_{n}(t)$ and 
$\beta_{n}(t)$ so as to annul $\lambda_{n}(t)$ and $\mu_{n}(t)$ and the 
matrix $M_{n}$ so as to take $A_{n}$ into its canonical form.

For $\tau =0$ the standardized form of $\hat X$ depends on the rank of the 
matrix~$A_{n}$. For $ \mbox{rank} A_{n}=2$ we can again transform 
$\lambda_{n}$ and $\mu_{n}$ into $\lambda_{n}=\mu_{n}=0$ and take $A_{n}$ 
into one of $3$ canonical forms. For $ \mbox{rank} A_{n}=1$ only one of the 
functions $\lambda_{n}$ or $\mu_{n}$ can be annulled. We choose it to be 
$\lambda_{n}(t)=0$. Then $A_{n}$ can be taken into one of the two standard 
matrices of rank~1 in $\mathbb R^{2 \times 2}$. For
 $ \mbox{rank} A_{n}=0$ both 
$\lambda_{n}(t)$ and $\mu_{n}(t)$ survive.

We thus obtain $10$ mutually inequivalent one dimensional symmetry algebras, 
listed below. The statement now is that any single vector field $\hat X$ of 
the form (\ref{symm.generator}) can be transformed by an allowed 
transformation into precisely one of these vector fields.

The next step is to determine the interactions for which a one dimensional 
symmetry group exists. To do this, we run through the canonical vector 
fields just obtained, substitute the corresponding \mbox{$\tau\,
 (=1 \;\mbox{or}\; 0)$}, 
$A_{n}\,,\,\lambda_{n}(t)$ and $\mu_{n}(t)$ into eqs. (\ref{remaining1}) and 
(\ref{remaining2}) and solve these equations for $F_{n}$ and $G_{n}$.

Following this procedure, we arrive at the following list of interactions 
and their one dimensional symmetry algebras.

\begin{alignat*}{2}
&A_{1,1}&\qquad\qquad &\hat X = \dt + a_n u_n \dun + d_n v_n \dvn,   \\
& & & F_{n}= e^{a_n t}f_n(\xi_k,\eta_k)\,,\\ 
& & &  G_n= e^{d_n t} g_n(\xi_k,\eta_k), \\
 & & &\xi_{k}= u_k e^{-a_k t},\quad \quad \eta_k =  v_k e^{-d_k t},  \\
 & & &k= n-1, n,n+1.  \\[6mm]
&A_{1,2}&\qquad\qquad &\hat X = \dt + (a_n u_n + v_n) \dun + a_n v_n \dvn,  \\
 & & &F_{n}=  e^{a_n t}\left[f_n(\xi_k,\eta_k) 
+ t\, g_n(\xi_k,\eta_k)\right]\,, \\
 & & &G_n= e^{a_n t}\,  g_n(\xi_k,\eta_k),      \\
 & & &\xi_{k}=  (u_k-t v_k) e^{-a_k t} ,  \quad\quad 
\eta_k = v_k e^{-a_k t}  ,    \\
 & & &k= n-1, n,n+1.  \\[6mm]
&A_{1,3}&\qquad\qquad &\hat X = \dt + (a_n u_n + b_n v_n)\dun +
 (-b_n u_n + a_n v_n)\dvn,\quad b_n > 0,  \\
 & & &\vect{F_{n}}{G_n}=e^{a_n t}\mat{\cos b_n t}{\sin b_n t}{-\sin b_n t}
{\cos b_n t} \vect{f_n(\xi_k,\eta_k)}{g_n(\xi_k,\eta_k)}        \\
 & & &\xi_{k}= r_k e^{-a_k t} , \quad\quad \eta_k = \theta_k + b_k t  , 
  \\
 & & &u_k= r_k \cos \theta_k, \quad\quad v_k=r_k \sin \theta_k,  \\
 & & &k= n-1, n,n+1. \\[6mm]
&A_{1,4}&\qquad\qquad &\hat X  = a_n u_n \dun + d_n v_n \dvn,\quad
\quad |a_n|\geq|d_n|,  \\
 & & &F_{n}= u_n f_n(\xi_\alpha,\eta_k,t) \, , \\ 
 & & & G_n=  v_n \,g_n(\xi_\alpha,\eta_k,t),       \\
 & & &\xi_{\alpha}= u_\alpha^{a_n}u_n^{-a_\alpha}, \quad\quad
 \eta_k =   v_k^{a_n}u_n^{-d_k},    \\
 & & &k= n-1, n,n+1,\quad\quad \alpha=n-1,n+1.  \\[6mm]
&A_{1,5}&\qquad\qquad&\hat X = (a_n u_n +v_n)\dun + a_n v_n \dvn  
,\quad\quad a_n \neq0, \\
 & & & F_{n}= v_n f_n(\eta_{\alpha},\xi_k,t) + v_n \ln{(v_n)}\, 
g_n(\eta_\alpha,\xi_k,t) \,, \\
 & & &  G_n=  a_n v_n  g_n(\eta_\alpha,\xi_k,t),  \\
 & & &\xi_{k}= a_k\frac{u_k}{v_{k}} -\ln{(v_k)} ,\quad\quad
 \eta_\alpha = v_\alpha^{a_n} v_n^{-a_\alpha} ,     \\
 & & &k= n-1, n,n+1,\quad\quad \alpha=n-1,n+1. \\[6mm]
\end{alignat*}


\begin{alignat*}{2}
&A_{1,6}&\qquad\qquad&\hat X = v_n\dun , \\
 & & & F_{n}= f_n(v_k,\xi_\alpha,t) + u_n \,g_n(v_k,\xi_\alpha,t), \\ 
 & & & G_n=  v_n \,  g_n(v_k,\xi_\alpha,t),  \\
 & & &\xi_{\alpha}= -v_\alpha u_n + v_n u_\alpha , \\
 & & &k= n-1, n,n+1,\quad\quad \alpha=n-1,n+1. \\[6mm]
&A_{1,7}&\qquad\qquad &\hat X =  (a_n u_n + b_n v_n)\dun +
 (-b_n u_n + a_n v_n)\dvn,\quad\quad b_n > 0,  \\
 & & &\vect{F_{n}}{G_n}=e^{-\frac{ a_n}{b_n}\theta_n}\mat{\cos \theta_n }
{-\sin \theta_n}{\sin \theta_n}{\cos \theta_n} 
\vect{f_n(\xi_k,\eta_\alpha,t)}{g_n(\xi_k,\eta_\alpha,t)}\\
 & & &\xi_{k}= r_k^{b_n} e^{a_k \theta_n}  ,\quad\quad 
\eta_\alpha = b_n \theta_\alpha - b_\alpha \theta_n   , 
  \\
& & &u_k= r_k \cos \theta_k ,\quad\quad v_k=r_k \sin \theta_k , \\
 & & & k= n-1, n,n+1,\quad\quad \alpha=n-1,n+1. \\[6mm]
&A_{1,8}&\qquad\qquad&\hat X = a_n u_n\dun +\mu_n(t) \dvn,\quad
\quad \mu_n \neq 0,  \\
 & & & F_{n}= u_n f_n(\eta_\alpha,\xi_k,t)  ,\\ 
 & & & G_n=  \frac{\ddot{\mu}_n}{\mu_n} \,v_n + g_n(\eta_{\alpha},\xi_k,t), \\
 & & &\eta_{\alpha}= \mu_n v_{\alpha}  - \mu_\alpha v_n ,\quad\quad
\xi_k = u_k e^{-\frac{a_k v_n}{\mu_n}},  \\
 & & &k= n-1, n,n+1,\quad\quad \alpha=n-1,n+1. \\[6mm]
&A_{1,9}&\qquad\qquad&\hat X = v_n\dun +\mu_n(t) \dvn,\quad
\quad \mu_n \neq 0,  \\
 & & & F_{n}= \frac{1}{2}\,\frac{\ddot{\mu}_n}{{\mu_n}^2}  v_n^2+
 v_n g_n(\eta_\alpha,\eta_n,\xi_\alpha,t) + 
f_n(\eta_\alpha,\eta_n,\xi_\alpha,t),\\ 
& & &  G_n=  \frac{\ddot{\mu_n}}{\mu_n} \,v_n +
 \mu_n  g_n(\eta_\alpha,\eta_n,\xi_\alpha,t), \\
 & & &\eta_{\alpha}= \mu_n^2 u_{\alpha} +\frac{1}{2}\, \mu_\alpha v_n^2 -
\mu_n v_n v_\alpha  
, \quad\quad
\xi_\alpha = \mu_\alpha v_n -\mu_n v_\alpha,  \\
 & & &\eta_n= \mu_n u_n -\frac{1}{2} \,v_n^2,\quad\quad \alpha=n-1,n+1. \\[6mm]
&A_{1,10}&\qquad\qquad&\hat X = \lambda_n(t)\dun +\mu_n(t)
\dvn,\quad
\quad \lambda_n,\;\;\mu_n \neq 0 , \\
 & & & F_{n}= \frac{\ddot{\lambda}_n}{\lambda_n}u_n + 
f_n(\eta_k,\xi_\alpha,t),\\
 & & & G_n=  \frac{\ddot{\mu_n}}{\mu_n} u_n +
  g_n(\eta_k,\xi_\alpha,t), \\
 & & &\xi_\alpha= \lambda_n u_{\alpha} - \lambda_\alpha u_n  
 ,\quad\quad
\eta_k = \mu_k u_n -\lambda_n v_k,  \\
 & & &k=n-1,n,n+1,\quad\quad \alpha=n-1,n+1. \\[6mm]
\end{alignat*}

We mention that the variables $\xi_k$ and $\eta_k$ are to be taken 
exactly as above. For instance $\xi_{n+1}$ is not an upshift of $\xi_n$.

The above results are summed up quite simply. Namely, the existence of a 
one dimensional symmetry algebra restricts the interaction terms $F_n$ and
 $G_n$ to two arbitrary functions of $6$ variables, rather than the original
 $7$ variables. The algebras $A_{1,1}$, $A_{1,2}$ and $A_{1,3}$ involve time
 translations. Hence, the time dependence in these cases is restricted: $F_n$
and $G_n$ depend on time explicitly and via invariant variables $\xi_k$ and 
$\eta_k$ that in turn depend explicitly on $t$.
 The algebras $A_{1,4},\ldots,A_{1,10}$ correspond to gauge transformations: 
the group transformations act on dependent variables only. The time variable
figures in the arbitrary functions. 

\section{HIGHER DIMENSIONAL SYMMETRY ALGEBRAS}\label{s4}

\subsection{General strategy}\label{41}

The commutator of two symmetry operators (\ref{symm.generator}) is an
 operator $X_3 = \left[X_1,X_2 \right]$ of the same form, satisfying
\begin{equation} \label{commutator}
\begin{array}{l}
 \tau_3= \tau_1\,\dot \tau_2- \tau_2 \dot\tau_1, \\[3mm] 
 A_{n,3} = -\left[ A_{n,1},A_{n,2} \right],  \\[3mm] 
 \vect{\lambda_{n,3}}{\mu_{n,3}} = \tau_1 \vect{ \dot \lambda_{n,2}}
{\dot \mu_{n,2}} - \tau_2 \vect{ \dot \lambda_{n,1}}
{\dot \mu_{n,1}} \\[3mm] 
 \qquad \qquad\; - \left( A_{n,1} + \frac{\ds \dot \tau_1}{2} \right)
 \vect{ \lambda_{n,2}}{\mu_{n,2}}  +  \left( A_{n,2} 
+ \frac{\ds \dot \tau_2}{2} \right)
\vect{ \lambda_{n,1}}{\mu_{n,1}}.   
\end{array}
\end{equation}

To obtain a finite dimensional Lie algebra of symmetry operators we see that
 the ``differential components''  $\tau_i(t)\dt$ must form a Lie algebra $L_d$,
 the ``matrix components'' $A_{n,i}$ must also form a Lie algebra $L_m$,
 homomorphic to $L_d$. Moreover, eq.(\ref{commutator}) shows that the
 ``functional components'' $(\lambda_{n,i}(t),\mu_{n,i}(t)\,)$ must satisfy
 certain cohomology conditions.

The algebra of diffeomorphisms of $\mathbb R^1$, $\left\{ \tau(t)\dt \right\}$ 
has only $3$ mutually nondiffeomorphic finite dimensional subalgebras, 
namely ${\mbox sl}(2,\mathbb R)$ and its subalgebras, realised e.g. as
\begin{equation}
\left\{ \dt,t\dt,t^2 \dt \right\}\;,\;\left\{ \dt,t\dt \right\}\;,
\;\mbox{and} \left\{ \dt \right\},
\end{equation}
respectively.

For $n$ fixed the matrices $A_n$ generate the Lie algebra of
 $gl(2, \mathbb R)$.
However, since the dependence on $n$ is arbitrary, an unlimited number of
 copies of $gl(2, \mathbb R)$ and its subalgebras is available.

We shall not perform a complete classification of possible symmetry algebras
 here. Instead, we shall first concentrate on ${\mbox sl}(2, \mathbb R)$
 symmetry algebras and show that, up to allowed transformations,
 four different  ${\mbox sl}(2, \mathbb R)$ symmetry algebras can 
be constructed. 
We then consider just one of these four and study its extensions to higher
 dimensional Lie algebras.

\subsection{Equivalence classes of ${\mbox sl}(2,\mathbb R)$ symmetry algebras}

Since ${\mbox sl}(2,\mathbb R)$ is a simple Lie algebra, it has no ideals. 
Hence a homomorphism between  ${\mbox sl}(2,\mathbb R)$ algebras is either an
 isomorphism, or one of the algebras is mapped onto zero.
Correspondingly we have $3$ possibilities to explore, we shall call them 
${\mbox sl}(2,\mathbb R)_d$, ${\mbox sl}(2,\mathbb R)_m$ and ${\mbox sl}
(2,\mathbb R)_c$ 
(where $d$ stands for ``differential'', $m$ for ``matrix'' and $c$ 
for ``combined'').

\begin{center}
\textbf{1.   The algebra ${\mbox sl}(2,\mathbb R)_d$}
\end{center}

 We have a priori

\begin{equation}
\begin{array}{l}
X_1 = \dt + \lambda_n(t) \dun +\mu_n(t) \dvn,  \\[2mm]
X_2 = t \dt + (\frac{1}{2} u_n + \rho_n(t))\dun + 
(\frac{1}{2} v_n + \sigma_n(t))\dvn,  \\[2mm]
X_3 =t^2 \dt + (t u_n + \omega_n(t))\dun + 
(t v_n + \kappa_n(t))\dvn .
\end{array}
\end{equation}
Using allowed transformations we transform $\lambda_n \rightarrow 0\,,
\,\mu_n \rightarrow 0$. The commutation relation $\left[X_1,X_2 \right]=X_1$ 
then requires $\dot \rho_n = \dot \sigma_n = 0$. A further allowed
 transformation (\ref{allowed}) with $ \tilde t (t) =t$, $M_n= I$ 
and $(\alpha_n,\beta_n)$ constant will not change $X_1$, but take 
$\rho_n \rightarrow 0$, $\sigma_n \rightarrow 0$ 
(while leaving $\lambda_n = \mu_n = 0)$.
The commutation relations  $\left[X_2,X_3 \right]=X_3$ and
  $\left[X_1,X_3 \right]=2 X_2$ then imply $\omega_n = \kappa_n = 0$.

\begin{center}
\textbf{2.  The algebra ${\mbox sl}(2,\mathbb R)_m$ }
\end{center}

A priori we have

\begin{equation}
\begin{array}{l}
X_1 =  b_n v_n \dun + \lambda_n(t) \dun + \mu_n(t) \dvn , \\[2mm]
X_2 =  a_n( u_n  \dun - v_n \dvn) + \rho_n(t)\dun + \sigma_n(t)\dvn,\\[2mm] 
X_3 =  c_n  u_n \dvn + \omega_n(t)\dun  + \kappa_n(t)\dvn .
\end{array}
\end{equation}
The structure constants cannot depend on $n$, so the commutation relations
 imply
\begin{equation}
a_n=a,\quad\quad b_nc_n = bc.
\end{equation}

Given that the product $b_n c_n$ does not depend on $n$, we can use an
 allowed transformation to take $b_n \rightarrow b$, $c_n\rightarrow c$. 
A further allowed transformation will take $\rho_n \rightarrow 0$,
 $\sigma_n \rightarrow 0$. The commutation relations then imply
 $\lambda_n=\mu_n =0$ and $\omega_n= \kappa_n =0$.

\begin{center}
\textbf{3. The combined algebra ${\mbox sl}(2,\mathbb R)_c$}
 \end{center}

In view of the above results we can write a ``combined'' algebra as

\begin{equation}
\begin{array}{l}
X_1 = \dt+ \alpha v_n \dun + \xi_n \dun + \eta_n \dvn,
\quad \alpha \neq 0,\\[2mm]
X_2 =  t \dt +\left[ (\frac{1}{2}+ \beta) u_n+\lambda_n \right] \dun +
 \left[ (\frac{1}{2}- \beta) v_n+\mu_n \right] \dvn, \\[2mm]
X_3 = t^2 \dt +\left( t u_n + \rho_n \right) \dun +
 \left( \gamma u_n + t v_n + \sigma_n \right) \dvn .
\end{array}
\end{equation}
We use allowed transformations to set $\alpha=1$, $\xi_n = \eta_n = 0$. The 
commutation relations then determine $\beta = \frac{1}{2}$, $ \gamma = -1 $.
The functions $\lambda_n(t)$, $\mu_n(t)$, $\rho_n(t)$ and $ \sigma_n(t)$ are
greatly restricted by the commutation relations. As a matter of fact, we
either have $\lambda_n= \mu_n= \rho_n = \sigma_n= 0$, or we can use  
allowed transformations to obtain $\lambda_n=t$, $\mu_n=1$, 
$ \rho_n = 2t^2$, $\sigma_n=2t$.
\newline \newline
We arrive at the following result.

\begin{theorem}
Precisely $4$ classes of ${\mbox sl}(2, \mathbb R)$ algebras can be realized by 
vector fields of the form (\ref{symm.generator}). 
Any such  ${\mbox sl}(2, \mathbb R)$ algebra can be taken by an allowed 
transformation
(\ref{allowed}) into precisely one of the following algebras:

\begin{alignat}{2}
&{\mbox sl}(2, \mathbb R)_1  \;:&\qquad& X_1=v_n \dun  \nonumber \\
& & & X_2 = \frac{1}{2}(u_n \dun - v_n \dvn) \label {sl2,1} \\
& & & X_3 = u_n \dvn   \nonumber  \\[6mm]
&{\mbox sl}(2, \mathbb R)_2  \;:&\qquad& X_1=\dt \nonumber \\
& & & X_2 = t \dt + \frac{1}{2}(u_n \dun + v_n \dvn)  \label{sl2,2} \\
& & & X_3 =t^2 \dt + t( u_n \dun +v_n \dvn)  \nonumber  \\[6mm]
&{\mbox sl}(2, \mathbb R)_3  \;:&\qquad& X_1=\dt + v_n \dun \nonumber \\
& & & X_2 = t \dt + u_n \dun  \label{sl2,3} \\
& & & X_3 =t^2 \dt + t u_n \dun + (t v_n -u_n )\dvn  \nonumber  \\[6mm]
&{\mbox sl}(2, \mathbb R)_4  \;:&\qquad& X_1=\dt + v_n \dun \nonumber \\
& & & X_2 = t \dt +( u_n +t)\dun + \dvn   \label{sl2,4}\\
& & & X_3 =t^2 \dt + (t u_n + 2 t^2)\dun + (t v_n -u_n +2t)\dvn. \nonumber  
\end{alignat}
\end{theorem}

\subsection{Indecomposable Lie algebras containing ${\mbox sl}(2, \mathbb
R)_1$}

A Lie algebra $L$ is called indecomposable if it cannot be written as a direct 
sum, $L= L_1 \oplus L_2$. A Lie algebra over $\mathbb R$ containing 
${\mbox sl}(2, \mathbb R)$ is either simple or it allows a nontrivial Levi 
decomposition [15]

\begin{equation}\label{Levi.decomp.}
L = S \triangleright R
\end{equation}
where $S$ is a semisimple Lie algebra and $R$ is the radical, that is the
maximal solvable ideal of $L$.

It follows from the results of Section \ref{41}
 that the only simple Lie algebras
 that can be constructed from operators of the form (\ref{symm.generator}) are 
the $4$ ${\mbox sl}(2, \mathbb R)$ algebras obtained in section $4.2$.
We can hence concentrate on Lie algebras of the form (\ref{Levi.decomp.}).

The algebra $S$ is either $sl(2, \mathbb R)_1$ itself, or the 
direct sum of 
${\mbox sl}(2, \mathbb R)_1$ with one or more other 
$ {\mbox sl}(2, \mathbb R)$ algebras.

Requiring that a symmetry operator $Y$ should commute with all elements of 
${\mbox sl}(2, \mathbb R)_1$ we find that $Y$ must have the form

\begin{equation}\label{central.elem.}
Y_0 = \tau \dt + \left( \frac{1}{2} \dot \tau + a_n \right) 
\left( u_n \dun + v_n \dvn \right).
\end{equation}
It is hence possible to construct precisely one semisimple Lie algebra properly
containing ${\mbox sl}(2, \mathbb R)_1$, namely the direct sum
\mbox{ ${\mbox sl}(2, \mathbb R)_1 \oplus {\mbox sl}(2, \mathbb R)_2$} with 
${\mbox sl}(2, \mathbb R)_2$
defined in eq.(\ref{sl2,2}).

Let us first introduce some notations for vector fields, to be used below. 
We put

\begin{alignat}{1}
& V(a_n) = a_n (u_n \dun + v_n \dvn)\,, \label{U} \\
& T(a_n) = \dt + a_n (u_n \dun + v_n \dvn) \,, \label{T}\\
& D(a_n) = t \dt + (1/2 +a_n) (u_n \dun + v_n \dvn) \,, \label{D}\\
& P(a_n) = t^2 \dt + (t +a_n) (u_n \dun + v_n \dvn) \,, \label{P}\\
& R(a_n) = (t^2+1) \dt +(t+ a_n) (u_n \dun + v_n \dvn) \,,\label{R} \\
& Y_u (\lambda_n) = \lambda_n(t) \dun, \quad\quad Y_v (\lambda_n) = 
\lambda_n(t) \dvn.
\label {YuYv}
\end{alignat}
In all cases we have $\dot{a}_n=0$, but $\lambda_n(t)$ can be a function
 of $t$.
 Both $a_n$ and $\lambda_n(t)$ can be functions of $n$.

Let us consider $S= {\mbox sl}(2, \mathbb R)_1$ and
 $S= {\mbox sl}(2, \mathbb R)_1 \oplus {\mbox sl}(2, \mathbb R)_2$ in
 eq.(\ref{Levi.decomp.})
separately.
\newline
\begin{center}
\textbf{A.  $\quad S= {\mbox sl}(2, \mathbb R)_1 $}
\end{center}

The considered Lie algebras will have a basis 
$\left\{ X_1,X_2,X_3, Y_1,\ldots,Y_n\right\}$ with $X_i$ given in
eq.(\ref{sl2,1}). The basis elements $\left\{ Y_1,\ldots,Y_n\right\}$ span
the radical $R$. The algebra $S$ acts on $R$ according to some linear,
 not necessarily irreducible, finite dimensional representation.

We start with the Cartan subalgebra $ \{X_2\} $ of ${\mbox sl}(2, \mathbb R)$.
It can be represented by a diagonal matrix in any 
finite dimensional representation.
Consider $Y \in R$. We have

\begin{equation}\label{cartan.elem.}
\left[X_2,Y\right]=p\,Y,
\end{equation}
 with $Y$ as in eq.(\ref{symm.generator}). Eq.(\ref{cartan.elem.}) implies

\begin{alignat}{3}
& &&\quad p\,\tau=0 \,,&&  \nonumber \\ 
& p\left(\frac{\dot \tau}{2} + a_n \right) =0 \,,&\qquad  
 &(p+\frac{1}{2}) \lambda_n =0\,,& \qquad & \left(p+1\right)b_n=0\,,
\label{cartan.cond.}\\
&  p\left(\frac{\dot \tau}{2} + d_n \right) =0\,,&\qquad   
&(p-\frac{1}{2})
 \mu_n =0 \,,& \qquad & (p-1)c_n=0 \,. \nonumber
\end{alignat}
For $p=0$ we obtain an operator that commutes not only with $X_2$,
 but with all of $ {\mbox sl}(2, \mathbb R)_1$, namely $Y_0$ of 
eq.(\ref{central.elem.}). This is
 a singlet representation of ${\mbox sl}(2,\mathbb R)$.
\newline
For $p=1$ , or $p= -1$ eq.(\ref{cartan.elem.}) forces $Y$ to be an element
of ${\mbox sl}(2,\mathbb R)_1$, in other words, no such $Y \in R$ exists.
\newline
For $p= \pm \frac{1}{2}$ we obtain $ Y_1 = \lambda_n(t)\, \dun$ and 
$Y_2 = \mu_n(t)\,\dvn$ respectively.
Acting with $X_1$ and $X_3$ on these operators, we find that the only 
representation of ${\mbox sl}(2,\mathbb R)_1$ that can be realized is a doublet
 one,
namely $\left\{ Y_u(\lambda_n),\,Y_v(\lambda_n) \right\}$ of 
eq.(\ref{YuYv}), with $\lambda_n(t)$ an arbitrary function of $n$ and $t$.
 The indecomposable Lie algebra 
$\left\{ X_1,X_2,X_3,Y_u(\lambda_n),Y_v(\lambda_n) \right\}$
is isomorphic to the special affine Lie algebra $\mbox{saff}(2,\mathbb R)$.

All further indecomposable symmetry algebras containing ${\mbox sl}
(2,\mathbb R)_1$
must be extensions of $\mbox{saff}(2,\mathbb R)$. The objects that we
 can add to
 $\mbox{saff}(2,\mathbb R)$ are either ${\mbox sl}(2,\mathbb R)$ doublets or
 singlets.
Let us run through all possibilities:
\newline

1.  We can add an arbitrary number $k$ of doublets of the form (\ref{YuYv})
where the $k$ functions 
$\left\{ \lambda_n^1(t),\lambda_n^2(t),\ldots, \lambda_n^k(t)\right\}$
must be linearly independent. However, we shall see in Section \ref{s5}
 that the
presence of $3$ such pairs forces the functions $F_n$ and $G_n$ in 
eq.(\ref{system}) to be linear.
Moreover, even two such pairs are compatible with a nonlinear interaction
only if they are of the form (or transformable into):
\begin{equation}
\lambda_n^1(t)=1,\qquad \lambda_n^2(t) = t\,.
\end{equation}

2.  We can add a singlet of the form (\ref{central.elem.}).
If we have $\tau=0$, then the commutation relations
$\left[Y_0,Y_u\right]$ and $\left[Y_0,Y_v\right]$ imply $a_n= a_{n+1}$
and we can set $a_n=1$. We obtain an affine Lie algebra 
$\mbox{gaff}(2, \mathbb R)_1$ consisting of $\mbox{saff}(2, \mathbb R)$ and
 $\mbox{V}(1)$ of eq.(\ref{U}).
\newline
If we have $\tau \neq 0$ in eq.(\ref{central.elem.}) and only one 
operator of this type, then we can use allowed tranformations to take 
$\tau(t)$ into $\tau(t)=1$. The commutation relations 
$\left[Y_0,Y_u\right]$ and $\left[Y_0,Y_v\right]$ then imply
\begin{displaymath}
\lambda_n(t) = R_n\, e^{(a_n+k)t},\quad \dot {R}_n =0.
\end{displaymath}
For $k=0$ the algebra is decomposable. For $k\neq 0$ we can use allowed 
transformations to put $k=-1$ and $R_n=1$ .
We obtain a second algebra isomorphic to $\mbox{gaff}(2, \mathbb R)$, 
but not conjugate to the previous one. We have
\begin{equation}\label{gaff2}
\mbox{gaff}(2, \mathbb R)_2 \sim \left\{ X_1,X_2,X_3,Y_u(e^{(a_n-1)t}),
Y_v(e^{(a_n-1)t}), T(a_n) \right\}.
\end{equation}
In the special case of $a_n=a_{n+1}$ in eq.(\ref{gaff2}) a further extension 
is possible.
We transform $\lambda = e^{(a-1)\,t}$ into $\lambda=1$, then $T(a_n)$ goes into
$D(b_n)$ with $ b_n=b_{n+1}\equiv b \neq -\frac{1}{2}$, since for 
$b=-\frac{1}{2}$ the algebra is decomposable.
\newline

3.  We can add $2$ singlets of the form (\ref{central.elem.}). If they commute,
 they must be $ \left\{ V(1),T(0) \right\}$. The obtained algebra is 
decomposable.
If they do not commute, they must form a two dimensional Lie algebra, namely
 $ \left\{ T(0),D(a)\;,\;a_n=a_{n+1}\equiv a \right\}$.
This implies $\lambda_n(t) \sim 1$, i.e. the entire radical is
$ \left\{ Y_u(1), Y_v(1),T(0),D(a) \right\}$ with $a\neq 1/2$ (the case
$a=1/2$ corresponds to a decomposable algebra).
\newline

4. If we add $3$ singlets, the only case corresponds to the radical 
$\{Y_u(1), Y_v(1), V(1),\,T(0),\,D(0)\}$. 
There will then be no invariant interaction 
(see below).
\newline

5.  Let us consider the special case of $2$ doublets of the form (\ref{YuYv}),
namely
\begin{equation}\label{2doublets}
Y_u (1)= \dun\;,\;Y_v(1) = \dvn\,, \quad Y_u(t) = t \dun\;,\; 
Y_v(t) =t \dvn.
\end{equation}
This algebra can be extended by a further element, namely
\begin{equation}
\begin{array}{c}
Z= \left( \tau_0 + \tau_1 \,t + \tau_2 \,t^2\right) \dt + 
(\frac{1}{2} \,\tau_1 + \tau_2 t + a ) 
\left( u_n \dun + v_n \dvn \right)\\[2mm]
 a_n=a_{n+1}\equiv a,
\end{array}
\end{equation}
where $\tau_0$, $\tau_1$ and $\tau_2$ are constants.
By allowed transformations we can take $Z$ into one of the $4$ operators 
$V(1)$, $T(a)$, $D(a)$ or $R(a)$ of (\ref{U}), (\ref{T}), (\ref{D})
and (\ref{R}), respectively.
\newline

6.   We can add a two dimensional algebra to (\ref{2doublets}), namely
\begin{displaymath}
\left\{ T(0),\,D(a) \right\},\;\; \left\{ T(0),\,V(1) \right\},
\;\;\left\{ V(1),\,D(0) \right\},\;\; \mbox{or} \;\;\left\{ V(1),\,R(0) \right\}.
\end{displaymath}

7. We can add only one three dimensional algebra to (\ref{2doublets}), namely
\begin{displaymath}
\left\{ T(0),\,D(0),\,V(1) \right\}.
\end{displaymath}

This completes the list of indecomposable symmetry algebras of the form 
(\ref{Levi.decomp.}) with $S={\mbox sl}(2, \mathbb R)_1$.
\newline
\begin{center}
\textbf{B.   $\quad S= {\mbox sl}(2, \mathbb R)_1 \oplus {\mbox sl}
(2, \mathbb R)_2$}
\end{center}

The algebra $S$ is itself decomposable. It gives rise to precisely two 
indecomposable symmetry algebra. First we have the one obtained by adding 
the Abelian ideal (\ref{2doublets}) to 
${\mbox sl}(2, \mathbb R)_1 \oplus {\mbox sl}(2, \mathbb R)_2$. Second,
we get an eleven dimensional algebra by adding $V(1)$ to the first case.

\subsection{Decomposable Lie algebras containing ${\mbox sl}(2, \mathbb R)_1$}

All decomposable Lie algebras $L_D$ can be obtained from the indecomposable 
$L_I$ ones, by adding their centralizers
\begin{equation}
L_D= L_I \oplus C\,, \qquad [C,L_I] =0.
\end{equation}

The centralizer $C$ must commute with all elements of 
${\mbox sl}(2, \mathbb R)_1$ 
and hence all of its elements will have the form of $Y_0$ of 
eq.(\ref{central.elem.}).   

Let us consider the individual indecomposable algebras $L_I$.
\newline
\newline
1.  $\quad L_I = {\mbox sl}(2, \mathbb R)_1$
\newline
\newline
The centralizer $C$ can be Abelian. Then we have the following possibilities: 
$
C = \left\{ V(a_{i,n})\;,\;i= 1,\ldots,k \right\}$ or
$C = \left\{ V(a_{i,n})\,,\,T(b_n)\;,\;i= 1,\ldots,k \right\}$.
The quantities $\ap,\ldots,a_{k,n}$ must form a set of $k$ linearly 
independent functions of $n$.
If the centralizer is non-Abelian, then we have either 
$C \sim {\mbox sl}(2, \mathbb R)_2$ or
$C= \left\{ T(0)\,,\,D(a)\;\right\}$. 
Both of these centralizers can be further extended by adding 
\mbox{$V(a_{i,n})\;,\;i=1,\ldots,k$,} (with $\ap,\ldots,a_{k,n}$ linearly
independent).
\newline
\newline
2. $\quad L_I = \mbox{saff}(2,\mathbb R)$
\newline
\newline
We must require $Y_0$ of eq.(\ref{central.elem.}) to commute with
$Y_u(\lambda_n)$ and $Y_v(\lambda_n)$ of eq.(\ref{YuYv}).
We obtain
\begin{equation} \label{4.26}
\lambda_n\,\left( \frac{1}{2}\,\dot \tau + a_n \right) - 
\tau\,\dot{\lambda}_n =0.
\end{equation} 
For $\tau =0$ eq.(\ref{4.26}) implies $\lambda_na_n =0$ and this is not 
allowed. For $\tau \neq 0$ we take $\tau \rightarrow 1$ by an allowed 
transformation, and eq.(\ref{4.26}) then implies 
$\lambda_n(t) = \gamma_n\, e^{a_n t}$.
A further allowed transformation will take $\gamma_n \rightarrow 1$.
We obtain the decomposable Lie algebra 
$ \mbox{saff}(2,\mathbb R) \oplus T(a_n)$.
In the special case $a_n=a_{n+1}$ we transform $\lambda_n(t) \rightarrow 1$
and obtain a larger centralizer, namely $\left\{ T(0),\,D(-\frac{1}{2})
\right\}$.
\newline
\newline
3. $\quad L_I = \mbox{gaff}(2,\mathbb R)_1$
\newline
\newline
A nontrivial centralizer exists only if we have $\lambda_n(t) =e^{\a t}$ in
$\mbox{saff}(2,\mathbb R)$. In
the case $\a \neq 0$, the centralizer is $C=\{T(\a)\}$. If $\a =0$ 
the centralizer is 
$C=\left\{ T(0),\,D(-\frac{1}{2}) \right\}$.
\newline
\newline
4. $\quad L_I = \mbox{gaff}(2,\mathbb R)_2$
\newline
\newline
The centralizer is $C= \left\{ T(a_n)-V(1) \right\}$. This algebra 
corresponds to the first one obtained in the case 
$L_I = \mbox{gaff}(2,\mathbb R)_1$ above. 

\subsection{Summary of possible symmetry algebras containing 
${\mbox sl}(2, \mathbb R)_1$}\label{45}

The classification of possible symmetry algebras can now be summed up rather 
simply. In addition to ${\mbox sl}(2, \mathbb R)_1$ of eq.(\ref{sl2,1}) 
we have a further algebra $L_C$ (the ``complementary'' algebra). 
The structure of each symmetry algebra is
\begin{equation}\label{complementary}
L= {\mbox sl}(2, \mathbb R)_1 \dot + L_C,\quad 
[{\mbox sl}(2, \mathbb R)_1,L_C]\subseteq L_C
,\quad [L_C,L_C] \subseteq L_C. 
\end{equation} 
The symbol $\dot +$ denotes a direct sum of vector spaces. Moreover, 
eq.(\ref{complementary}) shows that $L$ is either a direct sum or a 
semidirect one.
The algebra $L_C$ is also a representation space for 
${\mbox sl}(2, \mathbb R)_1$.
Irreducible representations in this case can be of dimension $1$ or $2$.
All higher dimensional representations are completely reducible into sums
of $1$ and $2$ dimensional representations.

For further use it is convenient to split the symmetry algebras into $4$ series,
 according to the structure of the Lie algebra $L_C$.
\begin{center}
\textbf{Series A}
\end{center}

$L_C$ is solvable and each element is an ${\mbox sl}(2, \mathbb R)_1$ singlet.
There exist $3$ different infinite dimensional Lie algebras of this type:
\begin{alignat}{2}
& A_1.& &\qquad \left\{ V(a_{k,n}) \right\} \\
& A_2.& &\qquad \left\{ T(b_n), V(a_{k,n})  \right\} \\
& A_3.& &\qquad\left\{T(0), D(b_n), V(a_{k,n}) \right\} 
\end{alignat}
In each case we have $k=1,2,\ldots$ and the expressions $a_k$ must be 
linearly independent funcions of $n$.
Taking $1 \leq k\leq N$ for some finite $N$, we obtain finite dimensional 
subalgebras.
\begin{center}
\textbf{Series B}
\end{center}

$L_C$ is solvable and contains precisely one 
${\mbox sl}(2, \mathbb R)_1$ doublet.
\begin{equation}
B_1= \left\{ Y_u(\lambda_n),Y_v(\lambda_n)\right\}.
\end{equation}
This is the indecomposable algebra $\mbox{saff}(2, \mathbb R)$ ($B_1$ 
together with ${\mbox sl}(2, \mathbb R)_1$). We have $\mbox{dim}L=5$.
\begin{equation}\label{B2}
B_2= \left\{ Y_u(\lambda_n),Y_v(\lambda_n), V(1)\right\}.
\end{equation}
$B_2$ is the indecomposable algebra $\mbox{gaff}(2, \mathbb R)_1$ with $\mbox{dim}L=6$.
\begin{equation}\label{B3}
B_3= \left\{ Y_u(e^{(a_n -1)t}),Y_v(e^{(a_n -1)t}), T(a_n)\right\}.
\end{equation}
$B_3$ is the Lie 
algebra $\mbox{gaff}(2, \mathbb R)_2$, isomorphic but not conjugate
 to $B_2$.
\begin{equation}
B_4= \left\{ Y_u(e^{a_n t}),Y_v(e^{a_n t}), T(a_n)\right\}.
\end{equation}
This algebra is $\mbox{saff}(2, \mathbb R) \oplus T(a_n)$.
\begin{equation}
B_5= \left\{ Y_u(1),Y_v(1), T(0), D(a)\right\}.
\end{equation}
The algebra $B_5$ is indecomposable (except if $a=-1/2$).
\begin{equation}\label{B6}
B_6= \left\{ Y_u(e^{(\a-1)t}),Y_v(e^{(\a-1)t}), T(\a), V(1)\right\}.
\end{equation}
The algebra $B_6$ is decomposable.
\begin{equation}\label{B7}
B_7= \left\{ Y_u(1),Y_v(1), T(0), D(0), V(1)\right\}.
\end{equation}
The algebra $B_7$ is indecomposable.
\begin{center}
\textbf{Series C}
\end{center}

$L_C$ contains two ${\mbox sl}(2, \mathbb R)$ doublets. 
The doublets could be characterized by any two functions $\lambda_{1,n}(t)$ 
and $\lambda_{2,n}(t)$.
However, we shall only be interested in the case $\lambda_1=1$, $\lambda_2=t$.
The others do not lead to invariant interactions.
Similarly, we do not need algebras containing $3$ or more doublets.
In all cases the algebra $L_C$ contains the elements (\ref{2doublets}).
For $\mbox{dim}\,L_C \geq 5$ it contains further elements.
We have
\begin{equation}
C_1 =\left\{ Y_u(1),Y_v(1), Y_u(t),Y_v(t)\right\}.
\end{equation}
Further we just list the additional elements
\begin{alignat}{2}
&C_2.& &\qquad \left\{ T(a) \right\}\,,\; a=0 \;\mbox{or}\;1 \\
&C_3.& &\qquad \left\{ D(a) \right\}\,,\;   \\
&C_4.& &\qquad \left\{ R(a) \right\}\,,\;  \\
&C_5.& &\qquad \left\{ V(1) \right\}\,,\;  \\
&C_6.& &\qquad \left\{ T(0), D(a) \right\}\,,\;  
\end{alignat}
In all cases above, $a$ does not depend on $n$ ($a_{n+1}= a_n$).
\begin{alignat}{2}
&C_7.& &\qquad \left\{ V(1),T(0) \right\}\,,\; \\
&C_8.& &\qquad \left\{ V(1),D(0) \right\}\,,\;   \\
&C_9.& &\qquad \left\{ V(1),R(0) \right\}\,,\;  \\
&C_{10}.& &\qquad \left\{ T(0), D(0), P(0) \right\}\sim 
{\mbox sl}(2,\mathbb R)_2\,.\; \\
&C_{11}.& &\qquad \left\{ T(0), D(0), V(1) \right\} \\
&C_{12}.& &\qquad \left\{ T(0), D(0), P(0), V(1) \right\} 
\end{alignat}
\begin{center}
\textbf{Series D}
\end{center}

$L_C$ contains  ${\mbox sl}(2,\mathbb R)_2$ and (possibly) further
 elements, namely
\begin{alignat}{2}
&D_1.& &\qquad \mbox{None,} \\
&D_2.& &\qquad \left\{V(\a)\right\}\,,\;   \\
&D_3.& &\qquad \left\{V(\ap),V(\as) \right\}\,,\;  \\
&D_4.& &\qquad \left\{Y_u(1),Y_v(1),Y_u(t),Y_v(t) \right\}\,\; \\
&D_5.& &\qquad \left\{Y_u(1),Y_v(1),Y_u(t),Y_v(t),V(1) \right\}\,\; 
\end{alignat}
($D_4$ coincides with $C_{10}$ and $D_5$ with $C_{12}$).


\section{THE INVARIANT INTERACTIONS}\label{s5}
\subsection{General Procedure and Interactions Invariant under $\mbox{SL}
(2,\mathbb{R})_1$}

In this section we shall find all interaction functions, invariant 
under symmetry groups, containing $\mbox{SL}(2,\mathbb{R})_1$. We make use 
of the subalgebra classification provided in Section \ref{s4}.

We first establish the form of the interaction, 
invariant under $\mbox{SL}(2,\mathbb{R})_1$ itself. To do this we set 
$\tau(t)=\lambda_n(t)=\mu_n(t)=0$ in the determining equations 
(\ref{remaining1}) and 
(\ref{remaining2}) and consider the equations obtained for 
$a_n=-d_n=1$, $b_n=c_n=0$, then 
$b_n=1$, $a_n=-d_n=c_n=0$, and finally $c_n=1$, $a_n=-d_n=b_n=0$. 
The general solution of the obtained system of $6$ equations 
can be
written in the following form:
\begin{equation}\label{intsl211}
\begin{array}{c}
\displaystyle{F_n=u_{n+1} f_n+u_n g_n,} \\ \\
\displaystyle{G_n=v_{n+1} f_n+v_n g_n,}
\end{array}
\end{equation}
where $f_n$ and $g_n$ are functions of $4$ variables each, 
namely
\begin{equation}\label{invariants}
\begin{array}{ll}
\displaystyle{t,\;\;\xi_n=u_{n+1} v_{n-1}-u_{n-1}v_{n+1},\;\;}\\ \\
\displaystyle{\xi_\alpha=u_{\alpha} v_{n}-u_{n}v_{\alpha},
\;\;\;\;\alpha=n\pm 1.}
\end{array}
\end{equation}
Note that $\xi_n$, $\xi_{n+1}$ and $\xi_{n-1}$ are as given in
eq.(\ref{invariants}). They are not upshifts or downshifts of each other.

We shall proceed further by dimension of the symmetry algebra and 
by its structure. Thus we can successively add ${\mbox sl}(2,\mathbb{R})$ 
singlets
of the form (\ref{central.elem.}) or doublets of the form 
(\ref{YuYv}). We continue 
adding 
symmetry elements, until the interaction is completely specified, 
i.e. involves no further arbitrary functions. We then solve the 
``inverse problem''. 
That is, we substitute the functions $F_n$ and $G_n$ back into the 
determining equations and solve for the symmetries. This provides a verification of previous 
calculations. More important, this procedure will find the 
largest symmetry algebra, allowed by any given interaction.

Obviously, all invariant interactions will have the form 
(\ref{intsl211}). 
It is the functions $f_n$ and $g_n$ that will be further refined 
and their dependence on the variables $\xi_k$ and $t$ will be restricted. 

For future convenience we write down two further forms of the
 $\mbox{SL}(2,\mathbb{R})_1$ invariant interaction functions,
 equivalent to (\ref{intsl211}).
 The first is
\begin{equation}
\begin{array}{c}\label{intsl212}
\displaystyle{F_n=u_{n+1}\frac{\xi_{n-1}}
{\xi_{n}}h_n+u_n k_n,} \\ \\
\displaystyle{G_n=v_{n+1}\frac{\xi_{n-1}}
{\xi_n}h_n+v_n k_n,}
\end{array}
\end{equation}
where $h_n$ and $k_n$ are arbitrary functions of the variables
 (\ref{invariants}). The second convenient form is
\begin{equation}\label{intsl213}
\begin{aligned}
F_n=&(\lambda_{n-1}u_{n+1}-\lambda_{n+1}u_{n-1})\phi_n 
+(\lambda_{n+1}u_n-\lambda_nu_{n+1})\psi_n\\ &
+\frac{\ddot{\lambda}_n}{\lambda_{n+1}}u_{n+1}, \\[2mm]
G_n=&(\lambda_{n-1}v_{n+1}-\lambda_{n+1}v_{n-1})\phi_n+
(\lambda_{n+1}v_n-\lambda_n v_{n+1})\psi_n\\ &
+\frac{\ddot{\lambda}_n}{\lambda_{n+1}}v_{n+1},
\end{aligned}
\end{equation}
where $\lambda_n(t)$ is some arbitrary function of $n$ and $t$ and 
$\phi_n$ and $\psi_n$ depend in an unspecified manner on
 the variables (\ref{invariants}).

\subsection{Interactions Invariant under Four dimensional Symmetry Groups}

As was shown in Section \ref{s4}, two types of $4$ dimensional symmetry 
algebras containing ${\mbox sl}(2,\mathbb{R})_1$ can exist. Both are 
decomposable according to the pattern $4=3+1$. Here and below 
we shall always list the operators that we can add to 
${\mbox sl}(2,\mathbb{R})_1$.

\begin{equation}
\hspace{-6cm}1.\;\;V(a_n)=a_n(u_n \partial_{u_n}+v_n\partial_{v_n})
\nonumber
\end{equation}

The invariant interactions will have the form (\ref{intsl212}), 
but $h_n$ and $k_n$ will depend on $3$ variables only.

\begin{itemize}
\item[(i)] $a_{n-1}+a_{n+1}\neq 0$. 

The variables are
\begin{equation}\label{tetas}
t,\;\;\eta_\alpha=(\xi_\alpha)^{a_{n-1}+\au}
(\xi_n)^{-a_n-a_\alpha},\;\;\alpha=n\pm 1. 
\end{equation}

\item[(ii)] $a_{n-1}+a_{n+1}= 0$.

The variables are
\begin{equation}\label{txieta}
t,\;\xi_n,\;\eta=(\xi_{n+1})^{a_{n+1}-a_n}(\xi_{n-1})^{a_{n+1}+a_n}.
\end{equation}
\end{itemize}

\begin{equation}
\hspace{-5.3cm}2.\;\;T(b_n)=\partial_t+b_n(u_n \partial_{u_n}+
v_n\partial_{v_n}).
\nonumber
\end{equation}

The invariant interaction will again have the form (\ref{intsl212}), 
however in this case $h_n$ and $k_n$ are arbitrary functions of the $3$
variables 
\begin{equation}\label{zeta}
\begin{array}{l}
\ds{\zeta_n=\xin e^{-(\bd+\bu)\,t},}\\ \\
\ds{\zeta_\alpha=\xi_\alpha
e^{-(\b+b_\alpha)\,t},\;\;\alpha=n\pm 1.}
\end{array}
\end{equation}

We see that adding further singlets of the type $V(\a)$ will restrict 
the variables in the functions $h_n$ and $k_n$, not however the
general form of eq.(\ref{intsl212}).

\subsection{Five dimensional Symmetry Groups}

  From the results of Section \ref{s4} we know that $3$ decomposable and $1$ 
indecomposable symmetry algebras of dimension $5$ can exist. 
Let us run through all
four possibilities.

\begin{center}
\bf{Decomposition $\bf{5=3+1+1}$}
\end{center}

\begin{equation}
\hspace{-1.1cm}1.\;\;V(a_{i,n})=a_{i,n}(u_n \partial_{u_n}+v_n\partial_{v_n}),
\;\;i=1,2,\quad a_{2,n}\neq\lambda \ap.
\nonumber
\end{equation}

The interaction is of the form (\ref{intsl212}). The functions $h_n$ and
$k_n$ depend on $2$ variables each, namely time $t$ and 
\begin{equation}\label{teta}
\eta=(\xid)^A(\xiu)^B(\xin)^C,
\end{equation}
\begin{equation}\label{ABC}
\begin{array}{l}
\ds{A=\ap(\asu+\asd)+\apu(\asd-\as)-\apd(\asu+\as),}\\ \\
\ds{B=-\ap(\asu+\asd)+\apu(\asd+\as)-\apd(\asu-\as),}\\ \\
\ds{C=\ap(\asu-\asd)-\apu(\asd+\as)+\apd(\asu+\as).}
\end{array}
\end{equation}

Note that the variable $\eta$ always exists since the condition 
\mbox{$A=B=C=0$} (and hence $\eta={\rm const}$) only occurs for 
$\apd \as-\ap \asd=0$, which implies
$\as=\lambda \ap$, $\lambda={\rm const}$, and this is not allowed.

\begin{equation*}
\hspace{-0cm}
2.\;\;V(\a)=\a(\u \dun+\v\dvn),\quad T(\b)=\dt+\b(\u \dun+\v\dvn).
\end{equation*}

The invariant interaction is as in eq.(\ref{intsl212}) with $h_n$ and 
$k_n$ functions of $2$ variables each. Namely: 
\begin{itemize}
\item[(i)] $\au+\ad \neq 0$
\begin{equation}\label{rho}
\rho_\alpha=(\zeta_\alpha)^{\au+\ad}(\zeta_n)^{-a_\alpha-\a},\;\;\alpha=n\pm 1
\end{equation}
with $\zeta_\alpha$, $\zeta_n$ as in eq.(\ref{zeta}).
\item[(ii)] $\au+\ad = 0$
\begin{equation}\label{xisigma}
\rho_n=\zeta_n,\;\;\sigma_n=(\zeta_{n-1})^{\au+\a}(\zeta_{n+1})^{\au-\a}.
\end{equation}
\end{itemize}

\begin{center}
\bf{Decomposition $\bf{5=3+2}$}
\end{center}

\begin{equation}
\hspace{-2.2cm}
3.\;\;T(0)=\dt,\;\;D(\b)=t\dt+\left(\frac{1}{2}+\b\right)(\u\dun+\v\dvn)
\nonumber
\end{equation}

We impose $\b \neq -\frac{1}{2}$, otherwise we have no invariant 
interaction.

We must distinguish $2$ subcases here.

\begin{itemize}
\item[(i)] $\bu+\bd+1\neq 0$

Interaction as in eq.(\ref{intsl212}) with
\begin{equation}\label{int32i}
\ds{h_n=(\xin)^{-\frac{2}{\bu+\bd+1}}p_n,}\quad
\ds{k_n=(\xin)^{-\frac{2}{\bu+\bd+1}}q_n},
\end{equation}
where $p_n$ and $q_n$ depend on $2$ variables, namely
\begin{equation}\label{chi}
\chi_\alpha=(\xi_\alpha)^{\bu+\bd+1}(\xin)^{-\b-b_\alpha-1},\;\;\alpha=n\pm 1.
\end{equation}

\item[(ii)] $\bu+\bd+1= 0,\;\;\bu+\b+1\neq 0$

\begin{equation}\label{int32ii}
\ds{h_n=(\xiu)^{-\frac{2}{\bu+\b+1}}p_n,}\quad
\ds{k_n=(\xiu)^{-\frac{2}{\bu+\b+1}}q_n},
\end{equation}
where $p_n$ and $q_n$ depend on
\begin{equation}\label{chixi}
\chi_n=(\xid)^{\bu+\b+1}(\xiu)^{-\bd-\b-1},\;\;\xin.
\end{equation}
\end{itemize}

Note that for
$\bu+\bd+1=0$, $\bu+\b+1=0$ we have $\b=-1/2$ and there is no invariant 
interaction.
\newline
\begin{center}
\bf{Indecomposable Lie algebra}
\end{center}

\begin{equation}
\hspace{-4cm}
4.\;\;Y_u(\lambda_n)=\lambda_n(t) \dun,\;\;Y_v(\lambda_n)=\lambda_n(t)\dvn.
\end{equation}

The invariant interaction is as in eq.(\ref{intsl213}), but the functions
$\phi_n$ and $\psi_n$ depend on only $2$ variables, namely
\begin{equation}\label{tomega}
t,\;\;\omega=\lambda_{n-1} \xiu-\lambda_n \xin-\lambda_{n+1}\xid.
\end{equation}


\subsection{Six dimensional Symmetry Groups}

\begin{center}
\bf{Decomposition $\bf{6=3+1+1+1}$}
\end{center}
\begin{equation}
\hspace{-4.1cm}1.\;\;V(a_{i,n})=a_{i,n}(u_n \partial_{u_n}+v_n\partial_{v_n}),
\;\;i=1,2,3. \nonumber
\end{equation}

The invariant interaction is as in eq.(\ref{intsl212}) but $h_n$ and $k_n$
are functions of $t$ only. We see that the coefficients $a_{i,n}$ do not
figure in the interaction. Hence, we can add an arbitrary number of
vector fields $V(a_{i,n})$, $i\in \mathbb{Z}$ to the symmetry algebra. In other
words, the symmetry algebra for the interaction (\ref{intsl212}) with $h_n$ 
and $k_n$ depending on $t$ alone is infinite dimensional.

\begin{equation*}
\hspace{-4cm}
\begin{array}{l}
\ds{2.\;\;V(a_{i,n})=a_{i,n}(\u \dun+\v\dvn),
\;\;i=1,2,}\\ \;\;\;\;\;
\ds{T(\b)=\dt+\b(\u \dun+\v\dvn).}
\end{array}
\end{equation*}

The invariant interaction is as in eq.(\ref{intsl212}) but $h_n$ and $k_n$
depend on $1$ variable only, namely
\begin{equation}\label{omegaM}
\omega=\eta {\mbox e}^{-2 t|M|},\;\;M=
\left(\begin{array}{lll}
\bd&\b&\bu\\ \apd&\ap&\apu\\ \asd&\as&\asu
\end{array}
\right)
\end{equation}
with $\eta$ as in eq.(\ref{teta}).

\pagebreak
\begin{center}
\bf{Decomposition $\bf{6=3+2+1}$}
\end{center}
\begin{equation*}
\hspace{-3.9cm}
\begin{array}{l}
\ds{3.\;\;V(a_{n})=a_{n}(\u \dun+\v\dvn),\;\;T(0)=\dt,}\\ 
\ds{\;\;\;\;\;D(\c)=t\dt+(\frac{1}{2}+\c)(\u \dun+\v\dvn).}
\end{array}
\end{equation*}

We start from eq.(\ref{intsl212}). The presence of $T(0)=\dt$ implies that
$h_n$ and $k_n$ do not depend on $t$. We first notice that if we have
\begin{equation}\label{gammas}
\gamma_n=\c+\frac{1}{2}=0\;\;\;\mbox{or}\;\;\;
\gamma_n=\c+\frac{1}{2}=\lambda \a
\end{equation}
then we must have $h_n=k_n=0$ (no invariant interaction). In all other cases,
invariance under $V(\a)$ and $D(\c)$ implies
\begin{equation}\label{int321}
\begin{array}{c}
\ds{h_n=(\xin)^\mu(\xiu)^\nu(\xid)^\rho p_n(\omega),}\quad
\ds{k_n=(\xin)^\mu(\xiu)^\nu(\xid)^\rho q_n(\omega),}\\ \\
\ds{\omega=(\xid)^A(\xiu)^B(\xin)^C}
\end{array}
\end{equation}
with $A$, $B$ and $C$ as in eq.(\ref{ABC}) with the substitutions
\begin{equation*}
\ap\rightarrow \c+\frac{1}{2},\quad\as\rightarrow \a.
\end{equation*}
\newline
The constants $\mu$, $\nu$ and $\rho$ in eq.(\ref{int321}) satisfy
\begin{equation}
\begin{array}{l}
\ds{(\au+\ad)\mu+(\au+\a)\nu+(\ad+\a)\rho =0,}\\ \\
\ds{(\gamma_{n+1}+\gamma_{n-1})\mu+(\gamma_{n+1}+\gamma_n)\nu+
(\gamma_{n-1}+\gamma_n)\rho=-2.}
\end{array}
\end{equation}
\newline
Thus, for $C\neq 0$ we can put 
\begin{equation*}
\mu= 0,\;\;\nu=2\,\frac{\a+\ad}{C},\;\;\rho=-2\,\frac{\a+\au}{C}.
\end{equation*}
For $C=0$, $A\neq 0$
\begin{equation*}
\mu=2\,\frac{\a+\au}{A},\;\;\nu=-2\,\frac{\au+\ad}{A},\;\;\rho=0.
\end{equation*}
For $C=A=0$, $B\neq 0$
\begin{equation*}
\mu=-2\,\frac{\ad+\a}{B},\;\;\nu=0,\;\;\rho=\,2\frac{\au+\ad}{B}.
\end{equation*}
The case $A=B=C=0$ corresponds to eq.(\ref{gammas}) and hence
to the absence of an invariant interaction.

\begin{center}
\bf{Decomposition $\bf{6=3+3}$}
\end{center}
\begin{equation}
\hspace{-7.4cm}4.\;\;{\mbox sl}(2,{\mathbb{R}})_1 
{\oplus} {\mbox sl}(2,{\mathbb{R}})_2 \nonumber
\end{equation}

The algebra ${\mbox sl}(2,\mathbb{R})_2$ is as in eq.(\ref{sl2,2}) and the 
invariant interaction is 
\begin{equation}
\label{int33}
\begin{array}{l}
\ds{F_n=\frac{1}{(\xin)^2}\left[\uu\frac{\xid}{\xin} p_n(\chi_{n+1},\chi_{n-1})
+\u q_n(\chi_{n+1},\chi_{n-1})\right],}\\ \\
\ds{G_n=\frac{1}{(\xin)^2}\left[\vu\frac{\xid}{\xin} p_n(\chi_{n+1},\chi_{n-1})
+\v q_n(\chi_{n+1},\chi_{n-1})\right],}
\end{array}
\end{equation}
$$
\chi_{n+1}=\frac{\xiu}{\xin},\;\;\chi_{n-1}=\frac{\xid}{\xin}.
$$

\begin{center}
\bf{Decomposition $\bf{6=5+1}$}
\end{center}
\begin{equation}
\hspace{-8.4cm}5.\;\;\mbox{saff}(2)\oplus A_1 \nonumber
\end{equation}

We have
\begin{equation*}
\begin{array}{l}
\ds{Y_u(e^{\a t})=e^{\a t}\dun,\;\;Y_v(e^{\a t})=e^{\a t}\dvn,}\quad
\ds{T(\a)=\dt+\a(\u\dun+\v\dvn).}
\end{array}
\end{equation*}
The invariant interaction will be as in eq.(\ref{intsl213}) 
with $\lambda_n=e^{\a t}$. The functions $\phi_n$ and $\psi_n$
will satisfy
\begin{equation}\label{int51}
\begin{array}{c}
\ds{\phi_n=e^{(\a-\ad-\au)\,t}K_n(\omega),}\quad
\ds{\psi_n=e^{-\au\, t}L_n(\omega),}\\ \\
\ds{\omega=e^{-(\a+\au)t}\xiu-e^{-(\au+\ad)t}\xin-e^{-(\ad+\a)t}\xid.}
\end{array}
\end{equation}

\begin{center}
\bf{Indecomposable symmetry algebras}
\end{center}
It was shown in Section \ref{s4} that two inequivalent 
$\mbox{gaff}(2)$ symmetry
 algebras
exist.

\begin{equation}
\hspace{-8.5cm}6.\;\;\mbox{gaff}(2,\mathbb{R})_1 \nonumber
\end{equation}

\begin{equation*}
\ds{Y_u(\lambda_n)=\lambda_n(t)\dun,\;\;
Y_v(\lambda_n)=\lambda_n(t)\dvn,\;\;V(1)=\u\dun+\v\dvn.}
\end{equation*}
The interaction is as in eq.(\ref{intsl213}), however $\phi_n$ and $\psi_n$ 
depend only on $t$. This means that the equations are linear and moreover
the equations (\ref{system}) for $u_n$ and $v_n$ are decoupled.

\begin{equation}
\hspace{-8.3cm}7.\;\;\;\mbox{gaff}(2,\mathbb{R})_2 \nonumber
\end{equation}

The algebra is as in eq.(\ref{gaff2}) (or (\ref{B3})), 
the interaction as in eq.(\ref{intsl213}) with
$\lambda_n(t)=e^{(\a-1)t}$. The functions $\phi_n$ and $\psi_n$ satisfy
\begin{equation}\label{intgaff2}
\begin{array}{l}
\phi_n=e^{-(\au+\ad-\a-1)t}K_n(\omega),\\ \\
\psi_n=e^{(-\au+1)t}L_n(\omega),
\end{array}
\end{equation}
$\omega$ as in eq.(\ref{int51}).

\subsection{Seven dimensional Symmetry Groups}

 \begin{center}
\bf{Decomposition $\bf{7=3+1+1+1+1}$}
\end{center}
We exclude the case
\begin{equation*}
V(a_{i,n})=a_{i,n}(\u\dun+\v\dvn),\;\;i=1,\ldots,4
\end{equation*}
since the only invariant interaction is (\ref{intsl212})
with $h_n$ and $k_n$ functions of $t$. We already know that the
symmetry algebra is infinite dimensional. 

\begin{equation*}
\hspace{-3.5cm}
\begin{array}{l}
\ds{1.\;\;V(a_{i,n})=a_{i,n}(\u \dun+\v\dvn),\;\;i=1,2,3,}\\ 
\ds{\;\;\;\;\;T(\b)=\dt+\b(\u \dun+\v\dvn).}
\end{array}
\end{equation*}

The interaction is as in eq.(\ref{intsl212}) with $h_n$ and $k_n$ constants
(depending on $n$). The algebra is actually infinite dimensional: we can take 
any number of operators $V(a_{i,n})$.

 \begin{center}
\bf{Decomposition $\bf{7=3+2+1+1}$}
\end{center}
\begin{equation*}
\hspace{-2.4cm}
\begin{array}{l}
\ds{2.\;\;V(a_{i,n})=a_{i,n}(\u \dun+\v\dvn),\;\;i=1,2,}\\ 
\ds{\;\;\;\;\;T(0)=\dt,\;\;D(\c)=t\dt+(\frac{1}{2}+\c)(\u \dun+\v\dvn).}
\end{array}
\end{equation*}

We put $\gamma_n=\c+\frac{1}{2}$. An invariant interaction exists if and
only if we have
\begin{equation}\label{delta}
\Delta =
\det{\left(
\begin{array}{ccc}
\gamma_n&\gamma_{n+1}&\gamma_{n-1}\\
\ap&\apu&\apd\\
\as&\asu&\asd
\end{array}
\right)}
\neq 0.
\end{equation}

The invariant interaction is that of eq.(\ref{intsl212}) with
\begin{equation}\label{perico}
h_n=\eta^kp_n,\;\;k_n=\eta^kq_n,\quad k=-\frac{2}{\Delta}.
\end{equation}
The variable $\eta$ is as in eq.(\ref{teta}), $p_n$ and $q_n$ are constants.
 \begin{center}
\bf{Decomposition $\bf{7=3+3+1}$}
\end{center}
\begin{equation}
\hspace{-6.4cm}3.\;\;{\mbox sl}(2,{\mathbb{R}})_1 
{\oplus} {\mbox sl}(2,{\mathbb{R}})_2{\oplus}A_1 \nonumber
\end{equation}

We have $A_1=\{V(\a)\}$. The invariant interaction can be obtained
from eq.(\ref{int33}). The additional invariance implied by the presence
of $V(\a)$ restricts $p_n$ and $q_n$ to
\begin{equation}\label{int331}
\begin{array}{l}
\ds{p_n=\left(\frac{\xiu}{\xin}\right)^{2\frac{\au+\ad}{\a-\ad}}
r_n(\omega),}\\
\\
\ds{q_n=\left(\frac{\xiu}{\xin}\right)^{2\frac{\au+\ad}{\a-\ad}}
s_n(\omega),}
\end{array}
\end{equation}
$$
\omega=(\xiu)^{\au-\a}(\xid)^{\a-\ad}(\xin)^{\ad-\au}
$$
and we must impose $\a\neq \ad$ (otherwise we have $F_n=G_n=0$).

 \begin{center}
\bf{Decomposition $\bf{7=6+1}$}
\end{center}
The algebra $\mbox{gaff}(2,\mathbb{R})_1$ does not allow any nonlinear 
interactions.
Let us consider $\mbox{gaff}(2,\mathbb{R})_2$ of eq.(\ref{gaff2}).

\begin{equation}
\hspace{-4.8cm}4.\;\;\mbox{gaff}(2,{\mathbb{R}})_2 
{\oplus}\{U=\u\dun+\v\dvn\} \nonumber
\end{equation}

The interaction is as in eq.(\ref{intsl213}) with $\phi_n$ and $\psi_n$
as in eq.(\ref{intgaff2}). Invariance under the dilations corresponding
to $U$ implies that $\phi_n$ and $\psi_n$ do not depend on $\omega$. Hence
the interaction is linear and decoupled.

 \begin{center}
\bf{Indecomposable Lie algebras}
\end{center}
\begin{equation*}
\hspace{-3.8cm}
\begin{array}{ll}
\ds{5.\;\;Y_u(\lambda_n)=\lambda_n(t)\dun,}&
\ds{Y_v(\lambda_n)=\lambda_n(t)\dvn,}\\
\ds{\;\;\;\;\;Y_u(\mu_n)=\mu_n(t)\dun,}&\ds{Y_v(\mu_n)=\mu_n(t)\dvn.}
\end{array}
\end{equation*}

We start from eq.(\ref{intsl213}) with $\phi_n$ and $\psi_n$ functions of $t$
 and $\omega$ as in eq.(\ref{tomega}). If $\phi_n$ and $\psi_n$ do not depend
 on $\omega$, the interaction is already linear and decoupled. Hence, $\omega$
must be invariant under the transformations corresponding to $Y_u(\mu_n)$ 
and $Y_v(\mu_n)$.
This implies that $\lambda_n$ and $\mu_n$ are independent of $n$. Further,
invariance implies
\begin{equation}\label{inutile}
\frac{\ddot{\lambda_n}}{\lambda_n}=\frac{\ddot{\mu_n}}{\mu_n}=\tilde{k}
\end{equation}
with $\tilde{k}=\mbox {const.}$ Eq.(\ref{inutile}) allows solutions
\begin{equation}
\vect{\lambda_n}{\mu_n}=\vect{\sin{kt}}{\cos{kt}},\;\;
\vect{{\rm sinh}kt}{{\rm cosh}kt},\;\;\vect{1}{t}.
\end{equation}
These solutions are all equivalent under allowed transformations.
We choose $\lambda_n=1$, $\mu_n=t$, i.e.
\begin{equation}\label{pavel}
Y_u(1)=\dun,\;\;Y_v(1)=\dvn,\;\;Y_u(t)=t\dun,\;\;Y_v(t)=t\dvn.
\end{equation}
The invariant interaction is
\begin{equation}\label{intind75}
\begin{array}{l}
\ds{F_n=(\uu-\ud)\phi_n(\omega,t)+(\u-\uu)\psi_n(\omega,t),} \\[3mm]
\ds{G_n=(\vu-\vd)\phi_n(\omega,t)+(\v-\vu)\psi_n(\omega,t)}
\end{array}
\end{equation}
with
\begin{equation}
\label{omsimple}
\omega=\xiu-\xid-\xin.
\end{equation}

\begin{equation*}
\hspace{-1cm}
\begin{array}{ll}
\ds{6.\;\;Y_u(1)=\dun,\;\;Y_v(1)=\dvn,\;\;T(0)=\dt,}\\ 
\ds{\;\;\;\;\;D(b)=t\dt+(\frac{1}{2}+b)(\u\dun+\v\dvn),\;\;
b\neq-\frac{1}{2},}\\ \ds{\;\;\;\;\;b=\mbox{const}.}&
\end{array}
\end{equation*}

The invariant interaction is as in eq.(\ref{intind75}) with
\begin{equation}\label{jaimito}
\phi_n=k_n\omega^{-\frac{2}{2b+1}},\;\;\psi_n=p_n\omega^{-\frac{2}{2b+1}}
\end{equation}
with $k_n$ and $p_n$ constants, $\omega$ as in eq.(\ref{omsimple}). 
For $b=-\frac{1}{2}$ there is no invariant interaction. For $b\neq 
-\frac{1}{2}$ the symmetry algebra is actually larger and includes 
$Y_u(t)=t\dun$ and $Y_v(t)=t\dvn$.

\subsection{Symmetry Groups of Dimensions $8$, $9$ and~$10$}

By now all invariant interactions have been specified up to arbitrary 
constants (depending on $n$), except those involving symmetry algebras
containing the subalgebra ${\mbox sl}(2,\mathbb{R})_1{\oplus}{\mbox sl}
(2,\mathbb{R})_2$,
or the subalgebra $\{Y_u(1),Y_v(1),Y_u(t),Y_v(t)\}$ of eq.(\ref{pavel}). 
Let us consider 
the remaining nonlinear interactions.

\begin{equation}
\hspace{-3.5cm}1.\;\;{\mbox sl}(2,\mathbb{R})_1{\oplus}{\mbox sl}(2,\mathbb{R})_2
{\oplus}\{V(\ap)\}{\oplus}\{V(\as)\} \nonumber
\end{equation}

The invariant interaction is obtained from eq.(\ref{int331}) by specifying 
$r_n(\omega)$ and $s_n(\omega)$ to be specific powers of $\omega$. The 
result is
\begin{equation}\label{intbiz}
\begin{array}{l}
\ds{F_n=\xin^{-2}\left[\uu\frac{\xid}{\xin}p_n+\u q_n\right]
(\xid)^{-\frac{2A}{D}}(\xiu)^{-\frac{2B}{D}}(\xin)^{2\frac{A+B}{D}},}\\ \\
\ds{G_n=\xin^{-2}\left[\vu\frac{\xid}{\xin}p_n+\v q_n\right]
(\xid)^{-\frac{2A}{D}}(\xiu)^{-\frac{2B}{D}}(\xin)^{2\frac{A+B}{D}}.}
\end{array}
\end{equation}
Here $p_n$ and $q_n$ are constants, $A$ and $B$ are as in eq.(\ref{ABC}) and
\begin{equation}\label{DD}
\begin{aligned}
D=&\ap(\asu-\asd)+\apu(\asd-\as)\\
&+\apd(\as-\asu).
\end{aligned}
\end{equation}
We assume $D\neq0$, otherwise there is no invariant interaction. 
In particular, we have $\ap\neq \apu$, $\as\neq \asu$.

\vspace{0.5cm}
\hspace{-0.1cm}$2.\;\;$ Algebras containing 
$(Y_u(1),Y_v(1),Y_u(t),Y_v(t))$
of
(\ref{pavel}) plus one additional operator $Z$. 
\vspace{0.5cm}

The interaction is as in eq.(\ref{intind75}) with a restriction on 
$\phi_n$ and $\psi_n$.
\begin{itemize}
\item[(i)]$Z=T(a)=\dt+a(\u\dun+\v\dvn),\;\; a\equiv\a=\au,$
\begin{equation}
\phi_n=\phi_n(\eta),\;\;\psi_n=\psi_n(\eta),\;\;\eta=\omega e^{-2a t}.
\end{equation}
\item[(ii)] $Z=D(a)=t\dt+(\frac{1}{2}+a)(\u\dun+\v\dvn),\;\; a\equiv\a=\au,$
\begin{equation}
\phi_n=\frac{1}{t^2}r_n(\eta),\;\;\psi_n=\frac{1}{t^2}s_n(\eta),
\;\;\eta=\omega t^{-(2a+1)}.
\end{equation}
\item[(iii)] $Z=R(b)=(t^2+1)\dt+(t+b)(\u\dun+\v\dvn),\;\; b\equiv\b=\bu,$
\begin{equation}
\phi_n=\frac{1}{(t^2+1)^2}r_n(\eta),\;\;
\psi_n=\frac{1}{(t^2+1)^2}s_n(\eta),
\end{equation}
$$\eta=\frac{\omega}{1+t^2}e^{-2b\,{\rm arctg}{t}}$$

with $\omega$ as in eq.(\ref{omsimple}) in all cases.

\item[(iv)] $Z=V(1)$. Then $\phi_n$ and $\psi_n$ depend only on $t$
and the interaction is linear.
\end{itemize}

We can add $2$ operators to those of eq.(\ref{pavel}) 
\begin{equation*}
T(0)=\dt,\;\;D(b)=t\dt+(\frac{1}{2}+b)(\u\dun+\v\dvn).
\end{equation*}
The invariant interaction coincides with that of eq.(\ref{jaimito}).

Finally, the interaction (\ref{intind75}) is invariant under a 
$10$ dimensional symmetry algebra of the form
\begin{equation*}
({\mbox sl}(2,\mathbb{R})_1\oplus {\mbox sl}(2,\mathbb{R})_2)
\triangleright\{Y_u(1),Y_v(1),Y_u(t),Y_v(t)\}
\end{equation*}
for
\begin{equation}
\phi_n=k_n\omega^{-2},\;\;\psi_n=p_n\omega^{-2},
\end{equation}
i.e. $b=0$ in eq.(\ref{jaimito}).


\section{Summary and Conclusions}\label{s6}

Let us first sum up the results on invariant interactions and the 
corresponding symmetry algebras.
We shall follow the summary of possible symmetry algebras outlined in 
Section \ref{45}. The results are presented in tables.

\textbf{Table 1.} The Series A of symmetry algebras.
The algebra $L_C$ of eq.(\ref{complementary}) consists entirely of 
$sl(2, \mathbb R)_1$ singlets. In the first column of Table $1$ we list the 
symmetry algebras. The number in brackets (e.g. $A_1(3)$) denotes the dimension
of the symmetry algebra. The notation for basis elements in column $2$ are
as in eq.(\ref{U})-(\ref{YuYv}). Note that if the functions $h_n$ and $k_n$
in the interaction (\ref{intsl212}) depend only on $t$ or are constants,
 then the
symmetry algebra is infinite dimensional, although the interaction is
 nonlinear.

The case $A_3(7)$ corresponds to an algebra $L$ with $\mbox{dim}\, L=7$ and
the interaction is completely specified. (see (\ref{intsl212}), 
(\ref{delta})-(\ref{perico}) ).
In other cases the functions $h_n$ and $k_n$ depend on $1$, $2$ or $3$ variables
 involving $u_k$ and $v_k$. 

\textbf{Table 2.} The Series B of symmetry algebras.
The symmetry algebras are either $5$ or $6$ dimensional. The interactions are as 
in eq.(\ref{intsl213}) and involve two arbitrary functions $\phi_n$ and $\psi_n$.
A B-type symmetry allows $\phi_n$ and $\psi_n$ to depend on just one variable
involving $u_k$ and $v_k$.
Any extension of the B type algebras will restrict $\lambda_n(t)$ to be
$\lambda_n=1$ and will involve a further pair with $\lambda_n=t$.
This takes us into the series C of symmetry algebras.

The algebras $B_2$, $B_6$ and $B_7$ of eq.(\ref{B2}), (\ref{B6}) and  
 (\ref{B7}) lead 
to linear interactions. Any interaction invariant with respect to 
$B_5$ will be invariant 
under a larger group, corresponding to a Lie algebra in the series C.
We do not include linear interactions in the tables and we list 
interactions together with their {\it maximal} symmetry 
algebras.

\textbf{Table 3.} The Series C of symmetry algebras.
The interaction will be as in eq.(\ref{intind75}) involving a variable $\omega$ as
in eq.(\ref{omsimple}). The algebras $C_5(8) $, $C_7(9)$, $C_8(9)$, $C_9(9)$,
$C_{11}(10)$, $C_{12}(11)$ absent in
the table, lead to a linear interaction.

For $C_6(9)$ and $C_{10}(10)$ 
the interactions are specified up to constants (that can depend on 
$n$). In all other cases, the arbitrary functions depend on one variable,
involving $u_k$ and $v_k$.

\textbf{Table 4.} The Series D of symmetry algebras.
There are $3$ such algebras of dimension $6$, $7$ and $8$, respectively.
They all lead to nontrivial invariant interactions of the form (\ref{int33}).
For $D_3(8)$ the interaction is completely specified.
We do not list $D_4(10)$ in Table $4$ since it coincides with $C_{10}(10)$ of
Table $3$. The algebra $D_5(11)$ corresponds to a linear interaction.

For each interaction we have verified that the given symmetry algebra is the 
maximal one.
\newline

A few words about the interpretation of the invariant interactions. From
eq.(\ref{intsl211}) and the variables (\ref{invariants}) we see that 
invariance under
$sl(2, \mathbb R)_1$ imposes very strong restrictions.
\newline
1. In particular, if the interaction is linear and $sl(2, \mathbb R)_1$
invariant, we must have
\begin{equation}
F_n = \sum^{n+1}_{k=n-1}A_k(t)u_k\,,\qquad G_n=\sum^{n+1}_{k=n-1}A_k(t)v_k,
\end{equation}
i.e. the equations (\ref{system}) for $u_k$ and $v_k$ decouple 
(into identical
equations for  $u_n$ and $v_n$ separately).
\newline
\newline
2. If the interaction terms $F_n$ and $G_n$ in eq.(\ref{intsl211}) are nonlinear, 
they always involve many-body forces.
That is, they cannot be written as sums of terms of the type $h_n(u_n,v_n)$
or $h_n(u_n,v_{n+1})$, etc\ldots
Indeed, each invariant variable $\xi_n$, $\xi_{n+1}$, $\xi_{n-1}$ itself
involves $4$ of the original variables $u_i$, $v_i$ simultaneously.
This many-body character becomes more pronounced when the invariance algebra
is larger.
\newline
\newline
3. The operators $V(a_n)$ correspond to site-depending dilations
\begin{equation}
\tilde{u}_n = e^{\epsilon a_n}\,u_n\,,\quad 
\tilde{v}_n = e^{\epsilon a_n}\,v_n\,.
\end{equation}
Invariance under two such one dimensional symmetry groups, generated by 
$\left\{V(a_{1,n}),V(a_{2,n}) \right\}$, where $a_{1,n}$ and $a_{2,n}$
are two linearly independent functions of $n$, introduces the 
symmetry variable
\begin{equation}\label{omD}
\omega_D \equiv (\xid)^A(\xiu)^B
(\xin)^C
\end{equation}
as in eq.(\ref{teta}). Here all $6$ variables are coupled together.
\newline
\newline
4.  The pair of operators $Y_u(\lambda_n)$, $Y_v(\lambda_n)$ induces 
site-dependent (and time-dependent) shifts of the dependent variables.
\begin{equation}
\tilde{u}_n= u_n + \epsilon \lambda_n(t)\,,\quad 
\tilde{v}_n= v_n + \epsilon \lambda_n(t).
\end{equation}
The corresponding invariant variable again involves all $6$ variables 
(see eq.(\ref{tomega})).
\begin{equation}\label{omT}
\omega_T\equiv \lambda_{n-1} \xiu
-\lambda_{n}\xin 
-\lambda_{n+1} \xid.
\end{equation}
A special case of the variable $\omega_T$ is obtained setting
$\lambda_n=\lambda_{n-1}=\lambda_{n+1} = 1$. This is the case of eq.(\ref{omsimple}), 
where
\begin{equation}\label{omS}
\omega=\omega_S=\xiu-\xin-\xid
\end{equation}
is invariant with respect to two such translations
\begin{equation}
\tilde{u}_n= u_n + \epsilon_1 +\epsilon_2 t\,,\quad 
\tilde{v}_n= v_n + \epsilon_1 +\epsilon_2 t.
\end{equation}
($\epsilon_1$ and $\epsilon_2$ are group parameters and hence constants).
\newline
\newline
A continuation of this study is in progress. It involves several aspects.
\newline 
The first is a study of the integrability properties of the equations 
that are
completely specified by their symmetries. These are, first of all, those 
with infinite dimensional symmetry groups, namely
\begin{equation}
\begin{aligned}
\ddot{u}_n &= u_{n+1}\frac{\ds\xid}
{\ds\xin}\, h_n + u_nk_n, \\[3mm]
 \ddot{v}_n &= v_{n+1}\frac{\ds \xid}
{\ds \xin}\, h_n + v_nk_n ,
\end{aligned}
\end{equation}
with $h_n$ and $k_n$ functions of $t$ or constants. (see $A_1(\infty)$ and
$A_2(\infty)$ in Table 1.)
\newline
\newline
Completely specified equations with finite dimensional symmetry algebras $L$
are the following ones:

\begin{itemize}
\item[(i)]
\begin{equation}\label{specified1}
\begin{aligned}
\ddot{u}_n&= \left( u_{n+1}\frac{\ds \xid}
{\ds \xin}\, p_n + u_nq_n \right) 
\omega_D^{-\frac{2}{\Delta}}, \\[3mm]
 \ddot{v}_n&= \left( v_{n+1}\frac{\ds \xid}
{\ds \xin}\, p_n + v_nq_n \right)\omega_D^{-\frac{2}{\Delta}},
\end{aligned}
\end{equation}
with $\omega_D$ as in eq.(\ref{omD}), $\Delta$ as in eq.(\ref{delta}). 
This is case $A_3(7)$ of Table 1.

\item[(ii)]
\begin{equation}\label{specified2}
\begin{aligned}
\ddot{u}_n&= \left[ (u_{n+1} -u_{n-1}) p_n + (u_n -u_{n+1})q_n
 \right]\omega_S^{-\frac{2}{2a+1}}, \\[2mm]
\ddot{v}_n&= \left[ ( v_{n+1} -v_{n-1}) p_n + (v_n -v_{n+1})q_n \right]\omega_S^{-\frac{2}{2a+1}},
\end{aligned}
\end{equation} 
with $\omega_S$ as in eq.(\ref{omS}), $p_n$, $q_n$, $a\neq -\frac{1}{2}$ constant.
This is case $C_6(9)$ of Table 3.

\item[(iii)]
For $a=0$ eq.(\ref{specified2}) is invariant under a $10$ dimensional
symmetry algebra, namely $C_{10}(10)$ of Table 3.

\item[(iv)]
\begin{equation} \nonumber
\begin{aligned}
\ddot{u}_n=  &(\xid)^{-\frac{2A}{k}}
 (\xiu)^{-\frac{2B}{D}}
 (\xin)^{2\frac{A+B-D}{D}} 
\left[  u_{n+1}\frac{\ds \xid}
{\ds \xin}\, p_n + u_nq_n  \right], \\[5mm]
\ddot{v}_n=&  (\xid)^{-\frac{2A}{D}}
 (\xiu)^{-\frac{2B}{D}}
 (\xin)^{2\frac{A+B-D}{D}}
\left[  v_{n+1}\frac{\ds \xid}
{\ds \xin}\, p_n + v_nq_n  \right],
\end{aligned}
\end{equation} 
with $p_n$ and $q_n$ depending only on $n$.
The constants $A$ and $B$ are given in eq.(\ref{ABC}), $D$ in 
eq.(\ref{DD}).
\end{itemize}
A further task is to complete the classification, that is to treat 
the cases
 of other $sl(2, \mathbb R)$ algebras and also of solvable symmetry 
algebras.

\begin{center}
\large{\bf ACKNOWLEDGEMENTS}
\end{center}
The authors thank D. Levi and M.A. Rodriguez for helpful
discussions. The research of 
S.L. and P.W. was partly supported by NSERC of Canada and FCAR du Qu\'ebec.
S.L. would like to thank the Departamento de
F\'{\i}sica Te\'orica II de la Universidad Complutense for 
their hospitality during his stay in Madrid. 
 D.G.U's work was partly
 supported by DGES grant PB95-0401. He would like 
to express his gratitude to 
the Centre de Recherches Math\'ematiques for their kind hospitality.

\clearpage
\begin{table}[h]
\hspace*{-0cm}
\caption{ Series $A$ of symmetry algebras. The interaction has the form
(\ref{intsl212}).}
\vspace{1cm}
\hspace*{-0cm}
\begin{tabular}[b]{cccc}
 & &Restrictions on & Variables and \\
No.& $L_C$ &$h_n$ and $k_n$&comments \\[1mm]
\hline
$A_1(3)$&-&-&$t,\;\;\xiu,\;\;\xid,\;\;\xin\;\;$(\ref{invariants})\\ \\
$A_1(4)$&$V(\a)$&-&$\left\{\begin{array}{c}\ds{t,\;\;\eta_{n+1},\;\;
\eta_{n-1}\;\;(\ref{tetas})}\\  \ds{t,\;\;\xin,\;\;\eta\;\;(\ref{txieta})}
\end{array}\right.$\\ \\
$A_1(5)$&$V(\ap),V(\as)$&-&$t,\;\;\eta\;\;(\ref{teta})$\\ \\
$A_1(\infty)$&$V(a_{i,n}),\;\;i\in\mathbb{Z}^{>}$&-&$t$\\[1mm]
 \hline \\
$A_2(4)$&$T(\b)$&-&$\zeta_{n+1},\;\;\zeta_{n-1},
\;\;\zeta_n\;\;(\ref{zeta})$\\ \\
$A_2(5)$&$T(\b),V(\a)$&-&$\left\{\begin{array}{c}\ds{\rho_{n-1},\;\;
\rho_{n+1}\;\;(\ref{rho})}\\  \ds{\rho_{n},\;\;\sigma_n\;\;(\ref{xisigma})}
\end{array}\right.$\\ \\
$A_2(6)$&$T(\b),V(\ap),V(\as)$&-&$\eta\;\;(\ref{omegaM})$\\ \\
$A_2(\infty)$&$T(\b),V(a_{k,n}),\;\;k\in\mathbb{Z}^{>}$&
$h_n$, $k_n$ constants&none \\ \hline \\
$A_3(5)$&$T(0),D(\b)$&(\ref{int32i}) or (\ref{int32ii})&
(\ref{chi}) or (\ref{chixi})\\ \\
$A_3(6)$&$T(0),D(\c),V(\a)$&(\ref{int321})&$\omega\;\;(\ref{int321})$\\ \\
$A_3(7)$&$T(0),D(\c),V(\ap),V(\as)$&(\ref{perico})&
none
\end{tabular}
\end{table}

\begin{table}[h]
\hspace*{-0.5cm}
\caption{Series B of symmetry algebras. The algebra includes one pair 
$Y_u(\lambda_n),\;\;Y_v(\lambda_n)$. The interaction has the form
(\ref{intsl213}).}
\vspace{.5cm}
\hspace*{-0.8cm}
\begin{tabular}[b]{cccc}
No.&Restrictions on $\lambda_n$, additional 
&Restrictions on $\phi_n$ and $\psi_n$ & Variables and  comments\\ 
&elements of $L_C$& & \\ 
\hline  
$B_1(5)$&-&-&$t,\;\;\omega$ as in (\ref{tomega})\\ \\
$B_4(6)$&$\lambda_n=e^{\a t},T(\a)$&(\ref{int51})&$\omega\;\;
(\ref{int51})$\\ \\
$B_3(6)$&$\lambda_n=e^{(\a-1)t},T(\a)$&(\ref{intgaff2})&
$\omega\;\;(\ref{int51})$
\end{tabular}
\end{table}

\begin{table}[h]
\hspace{-0.8cm}
\caption{Series C symmetry algebras. The algebras contain
 $sl(2,\mathbb{R})_1$ $Y_u(1),\;\;Y_v(1),\;\;Y_u(t),\;\;
Y_v(t)$ and possibly additional elements. The interaction is as in
eq.(\ref{intind75}).}
\vspace{.5cm}
\hspace*{-0.8cm}
\begin{tabular}[b]{cccc}
No&Additional elements
&Conditions on $\phi_n$ and $\psi_n$& Variables\\
\hline 
$C_1(7)$&-&-&$\omega,\;\;t$ (\ref{omsimple})\\ \\
$C_2(8)$&$T(a)$&-&$\eta =\omega e^{-2at}$\\ \\
$C_3(8)$&$D(a)$&$\phi_n=t^{-2}r_n(\eta),\;\;\psi_n=t^{-2}s_n(\eta)$&
$\eta=\omega t^{-(2a+1)}$\\ \\
$C_4(8)$&$R(b)$&$\phi_n=(t^2+1)^{-2}r_n(\eta),$&$\eta=\omega(t^2+1)^{-1}$\\
&&$\psi_n=(t^2+1)^{-2}s_n(\eta)$&$e^{-2b\,{\rm arctg}{t}}$\\ \\
$C_6(9)$&$T(0),D(a)$&$\phi_n=k_n\omega^{-\frac{2}{2a+1}},\;\;
\psi_n=p_n\omega^{-\frac{2}{2a+1}}$&none \\ 
&&$k_n,\;\;p_n$ constants, $2a+1\neq 0$&\\ \\
$C_{10}(10)$& $T(0),D(0),P(0)$&$\phi_n=k_n\omega^{-2},\;\;
\psi_n=p_n\omega^{-2}$& none
\end{tabular}
\end{table}

\begin{table}[h]
\hspace{-0.8cm}
\caption{Series D of symmetry algebras. The algebra contains 
$sl(2,\mathbb{R})_1{\oplus}\,sl(2,\mathbb{R})_2$. The interaction
has the form (\ref{int33}).}
\vspace{.5cm}
\hspace*{-0.8cm}
\begin{tabular}[b]{cccc}
No&Additional elements in $L_C$
&Conditions on $p_n$ and $q_n$& Variables\\
\hline
$D_1(6)$&-&-&$\chi_{n+1},\,\chi_{n-1}$ as in (\ref{int33})\\ \\
$D_2(7)$&$V(\a)$&(\ref{int331})&$\eta$ as in (\ref{int331})\\ \\
$D_3(8)$&$V(\ap),V(\as)$&(\ref{intbiz})&-\\ \\
\end{tabular}
\end{table}
\end{document}